                        \newif\ifboyscout                         
                        \boyscoutfalse 
                        \newif\ifpreparepdf                       
                        \preparepdftrue 
                        \newif\ifhighlightedits
                        \highlighteditsfalse
                        \newif\ifrevision
		                \revisionfalse
                        \newif\ifrefereeone
                        \refereeonefalse
                        \newif\ifrefereetwo
                        \refereetwofalse
                
        \ifboyscout
\documentclass[aps,preprint,showpacs,longbibliography,nofootinbib]{revtex4-1} 
        \else
\documentclass[aps,showpacs,superscriptaddress,longbibliography,nofootinbib]{revtex4-1} 
        \fi

\ifpreparepdf
    \usepackage{color} 
    \usepackage[colorlinks]{hyperref} 
\else 
\fi

\usepackage{graphicx,amssymb,amsbsy}
\usepackage{ifthen}
\usepackage[percent]{overpic}
\usepackage[colorlinks]{hyperref}
\usepackage{amsmath}
\usepackage{bm}

\graphicspath{{figs/}}  

    \ifboyscout 

\newcommand{\notetarget}[2]{\hypertarget{#1}{#2}}
    \else  
    \ifrevision

\newcommand{\notetarget}[2]{\hypertarget{#1}{#2}} 
    \else

\newcommand{\notetarget}[2]{\hypertarget{#1}{#2}}
    \fi
    \fi

\newcommand{\arXiv}[1]{\href{http://arXiv.org/abs/#1}{arXiv:#1}}

\newcommand{\rf}     [1] {~\cite{#1}}
\newcommand{\rfp}[1] {~\citep{#1}}
\newcommand{\refref} [1] {ref.~\cite{#1}}

\newcommand{\refrefs}[1] {refs.~\cite{#1}}
\newcommand{\refeq}  [1] {(\ref{#1})}
\newcommand{\reffig} [1] {fig.~\ref{#1}}
\newcommand{\refFig} [1] {Fig.~\ref{#1}}
\newcommand{\reffigs} [1] {figs.~\ref{#1}}
\newcommand{\refFigs} [1] {Figs.~\ref{#1}}

\newcommand{\beq}{\begin{equation}}

\newcommand{\eeq}{\end{equation}}
\newcommand{\ee}[1] {\label{#1} \end{equation}}
\newcommand{\bea}{\begin{eqnarray}}

\newcommand{\eea}{\end{eqnarray}}

\newcommand{\etal}{{\em et al.}}    

\newcommand{\statesp}{state space}

\newcommand{\zeit}{\ensuremath{t}}  

\newcommand{\bseq}{\begin{subequations}}
\newcommand{\eseq}{\end{subequations}}

\newcommand{\NSe}{Navier-Stokes equations}
\newcommand{\Reynolds}{\textit{Re}}  

\newcommand{\eigExp}[1][]{
     \ifthenelse{\equal{#1}{}}{\ensuremath{\lambda}}{\ensuremath{\lambda^{(#1)}}}}
\newcommand{\eigRe}[1][]{
     \ifthenelse{\equal{#1}{}}{\ensuremath{\mu}}{\ensuremath{\mu^{(#1)}}}}
\newcommand{\eigIm}[1][]{
     \ifthenelse{\equal{#1}{}}{\ensuremath{\omega}}{\ensuremath{\omega^{(#1)}}}}

\newcommand{\ssp}{\ensuremath{a}}                

\newcommand{\vfield}{\ensuremath{\bm{u}}}

\newcommand{\product}{\ensuremath{\mathcal{P}}} 
\newcommand{\dissip}{\ensuremath{\mathcal{D}}} 

\begin{document}

\title{
Upper edge of chaos and the energetics of transition in pipe flow
}
\author{Nazmi Burak Budanur}
\affiliation{Nonlinear Dynamics and Turbulence Group,
             IST Austria,
             3400 Klosterneuburg, Austria}
\author{Elena Marensi}
\affiliation{School of Mathematics and Statistics,
             University of Sheffield, S3\,7RH, UK}
\author{Ashley P. Willis}
\affiliation{School of Mathematics and Statistics,
             University of Sheffield, S3\,7RH, UK}
\author{Bj\"{o}rn Hof}
\affiliation{Nonlinear Dynamics and Turbulence Group,
             IST Austria,
             3400 Klosterneuburg, Austria}

\date{\today}

        \begin{abstract}
        In the past two decades, our understanding of the transition to 
        turbulence in shear flows with linearly stable laminar solutions has
        greatly improved. Regarding the susceptibility of the laminar 
        flow, two concepts have been particularly useful:
        the edge states and the minimal seeds. 
        In this nonlinear picture of the transition, 
        the basin boundary of turbulence is set by the 
        edge state's stable manifold and this manifold
        comes closest in energy to the laminar equilibrium at the minimal seed.
        We begin this paper by presenting numerical experiments in which
        three-dimensional perturbations are too energetic to 
        trigger turbulence in pipe flow but they do lead to turbulence
        when their amplitude is reduced. 
		We show that this seemingly counter-intuitive observation is 
		in fact consistent with the fully nonlinear description of the 
		transition mediated by the edge state. 
        In order to understand the physical mechanisms behind this process, 
        we measure the turbulent kinetic energy production and dissipation 
        rates as a function of the radial coordinate. 
        Our main observation is that the transition to turbulence relies on 
        the energy amplification away from the wall, as opposed to the 
        turbulence itself, whose energy is predominantly produced near the wall. 
        This observation is further supported by the similar analyses on the 
        minimal seeds and the edge states. 
        Furthermore, we show that the time-evolution of 
        production-over-dissipation curves provide a clear distinction between the 
        different initial amplification stages of the transition to turbulence from 
        the minimal seed.
        \end{abstract}

\maketitle

\section{Introduction}
\label{sec:intro}

Transition to turbulence in shear flows with linearly stable laminar 
solutions has baffled researchers since the seminal experiments of 
Reynolds\rf{R1883} for more than a century. Despite its ubiquity
 in the nature and applications, many aspects of 
this phenomenon have only recently been understood, in part 
thanks to the 
availability of new computational and experimental tools.  
In our new understanding of the transition without a linear instability,
 turbulence is triggered by 
finite-amplitude disturbances to the laminar flow that push the state 
across the so-called 
``edge of chaos''\rfp{SYE05}, which separates perturbations that 
can trigger turbulence from those that cannot. 
This description of transition follows a fully nonlinear 
geometrical approach to the problem, which adopts ideas from 
the dynamical systems theory. 

Fluid motion through a channel or a pipe can be thought of as 
a trajectory in the infinite-dimensional \statesp\ of 
the attainable velocity fields. This viewpoint of the fluid dynamics
was put forward by Hopf\rf{hopf48}, who also conjectured that the 
turbulence should asymptotically be confined in a 
finite-dimensional manifold in the \statesp\ due to the presence
of dissipation in the governing \NSe . While a rigorous proof of 
whether this indeed is the case is yet to be found, the assumption
of finite-dimensionality is tacit in all computational 
fluid dynamics, where the velocity fields are expressed on a finite 
grid or an equivalent spectral expansion. 
In fact, a finite-dimensional numerical representation constitutes an 
approximation to the infinite dimensional state space. If each 
numerical degree-of-freedom is assigned to an axis, then a snapshot 
of the fluid is a point in this high-dimensional space and its 
time-evolution under the Navier-Stokes equation is a trajectory. 

Once the dynamical systems viewpoint is established, the transition 
question becomes a geometrical one. Which initial points in the 
\statesp\ eventually connects to the turbulence? Or conversely, which 
initial conditions decay onto the laminar equilibrium as they are 
evolved in time? In low dimensional chaotic systems with 
coexisting attracting sets, the basin boundary between the attractors
is often found to be the stable manifold of a saddle-type invariant 
solution\rfp{ASY1997ch10,GOY1982,GOY1983}. In numerical studies
of such systems, the saddle-type solutions can be probed via a 
bisection algorithm that 
refines initial conditions between those that end up in either attractor, 
yielding an approximation of an initial condition on the stable 
manifold of the saddle solution. In the case of transition to 
turbulence without a linear instability of the laminar solution, 
any initial condition that is weak enough will decay. Thus, if an 
initial condition that yields turbulence is found, then this 
initial condition could be scaled down until it can no longer 
trigger the transition, in order to initiate a bisection algorithm. 
Through such a bisection search in a computational 
study of channel flow, Itano \& Toh\rf{IT01} found the basin boundary 
between laminar and turbulence in channel flow to be set by a traveling wave 
solution of the saddle-type. On a reduced-order model of a 
shear flow, Skufca \etal \rf{SYE05} showed that the asymptotic dynamics
on the laminar-turbulent boundary can be periodic, or even chaotic. 
Similar computational studies of shear 
flows\rfp{TI03,SchEckYor07,SGLDE08,MMSE09,ScMaEc10,ZamEck14a,KKSDEH2016}
yielded various types of asymptotic states in the laminar-turbulent 
boundary. 
In most of the current literature, 
irrespective of the type 
of their time-dependence, 
such asymptotic states at the laminar-turbulent boundary 
are referred to as the ``edge states''; and the bisection-based 
methods that probe edge states are called ``edge tracking''. 
In the pipe flow setting, which 
we are going to consider in the current study, the edge state appears to 
be chaotic with transient visits to traveling 
wave solutions\rfp{budanur-hof-2018}.

If, in the subcritical regime, only some initial disturbances 
of sufficiently large energy
 can trigger transition, then a key question, both from a practical and scientific viewpoint,
is the following.
What is the weakest 
perturbation that is strong enough to trigger turbulence? 
The practical importance of identifying such minimal perturbations
is due to the
fact that avoiding them would be 
crucial in settings where turbulence is undesirable. 
In the edge state picture, 
such a minimal perturbation is the one that touches the stable manifold
of the edge state at the point where the distance --in some norm--
between the manifold and the laminar solution is the shortest. 
Pringle \etal \rf{pringle-kerswell-2010} formulated a nonlinear 
optimization method for identifying ``minimal seeds'' for triggering 
turbulence by searching for perturbations to the laminar flow whose 
energies amplified the most in a finite time horizon. 
Pringle \etal \rf{pringle-kerswell-2010} and other studies\rfp{pringle-etal-2012,
	duguet-etal-2013,cherubini-palma-2014,kerswell-etal-2014}
demonstrated that the time-forward dynamics of minimal seeds indeed
approach the edge states, evidencing that the minimal seeds
lie on the edge states' stable manifold.
The flow structures of the minimal perturbations appeared
completely different from those that can be expected from weakly 
nonlinear theories, demonstrating the strongly nonlinear nature of
the transition problem.
The energy growth of the minimal seed was found to occur via three mechanisms, coupled together by the nonlinear effects to produce a larger overall growth than a linear optimal.
The initially fully localized streaks are slightly unwrapped and tilted away from the wall by the inviscid Orr process.
This is then followed by the oblique-wave mechanism in which the helical modes quickly grow transiently and feed energy into the streamwise-independent modes. In the last phase, the well-known lift-up process, the streamwise rolls experience non-normal energy growth and advect the shear to drive the streaks. The Orr and oblique-wave phases occur in a very short time scale (less than 4 advective units) and give rise to most of the growth experienced by the minimal seed. By comparing the energy of the minimal seed at different phases of its evolution with the critical energies of different randomized disturbances, Marensi \etal \rf{marensi-etal-2019} showed that the minimal seed, despite being quite `special', evolves to a structure more similar to a `natural' disturbance during the Orr and oblique-wave phases. In particular, the minimal seed at the end of the oblique-wave phase was found to be a reasonable proxy to characterize the critical initial energy of typical ambient disturbances, thus proving to be a useful tool to measure transition thresholds in realistic scenarios.

We begin this paper with numerical experiments that explore 
the logical opposite of the ``minimal seed'' question:
Are there three-dimensional\footnote{\notetarget{twodim}{Note} that this question is 
	only nontrivial for three-dimensional perturbations, 
	since two-dimensional axially-independent initial states in pipe flow
	always decay to laminar, regardless of their amplitude\rf{Joseph1971}.} 
perturbations to the laminar flow that are 
\textit{too strong} such that the dynamics uneventfully laminarize? 
Although counter-intuitive, such perturbations may exist given the
nonlinear nature of the transition. Furthermore, such 
perturbations may provide insights for control applications
in order to eliminate turbulence.
 In agreement with the recent 
experimental observations of K\"uhnen \etal \rf{kuhnen-etal-2018a}, we find that highly-turbulent 
initial conditions with flat axial velocity profiles could lead to 
a complete laminarization. Via edge-tracking, we show that these 
initial conditions are indeed on the laminarizing side of the 
edge state's stable manifold. Upon measuring turbulence energy 
production and dissipation on these initial conditions, we observe 
that the flattening of the velocity profile leads to a drastic 
decrease of turbulence production in the bulk region of the pipe.
We then measure the same quantities on edge states and minimal seeds
and conclude that the strong energy amplification away from the wall
is a characteristic property of the transition in pipe flow. 
Furthermore, we show that the curves of production over dissipation serve 
as diagnostic tools for identifying the Orr and oblique-wave phases
of the transition from the minimal seed. 

\section{The upper edge of chaos}
\label{s-upper}

Recent experiments and numerical simulations by K\"uhnen \etal \rf{kuhnen-etal-2018a} showed 
that flattening the flow 
profile by various control methods leads to the complete decay of 
turbulence for Reynolds numbers up to 40000. The main idea behind this was to disrupt the energy
production in such a way that the turbulence cannot sustain itself. 
K\"uhnen \etal \rf{kuhnen-etal-2018a} attributed this to the linear 
part of the Navier-Stokes equations and used transient growth around a 
mean profile as a proxy for measuring this effect. Indeed, they found
that the complete relaminarization is observed when the transient
growth of a mean profile was below a certain level. We begin our analysis 
here by studying similar events in the \statesp .

\begin{figure}
	\setlength{\unitlength}{0.45\textwidth}
	(a) \includegraphics[width=0.95\textwidth]{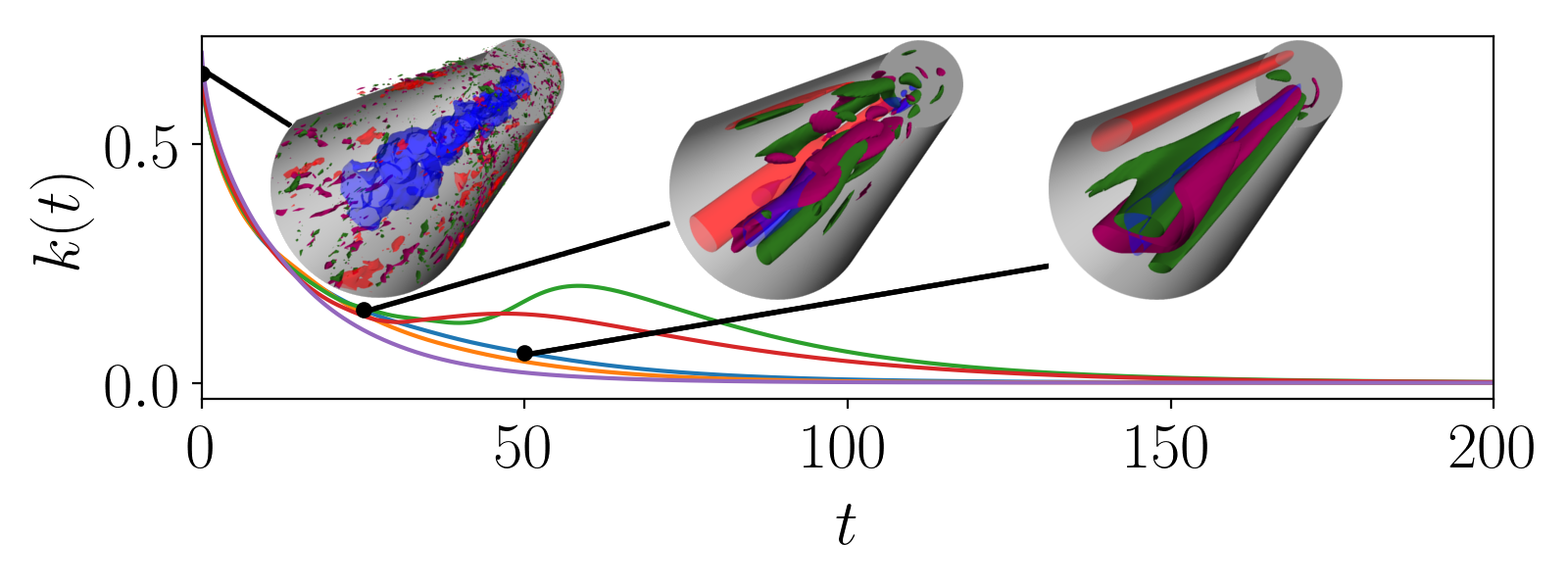} \\
	(b) \includegraphics[width=\unitlength]{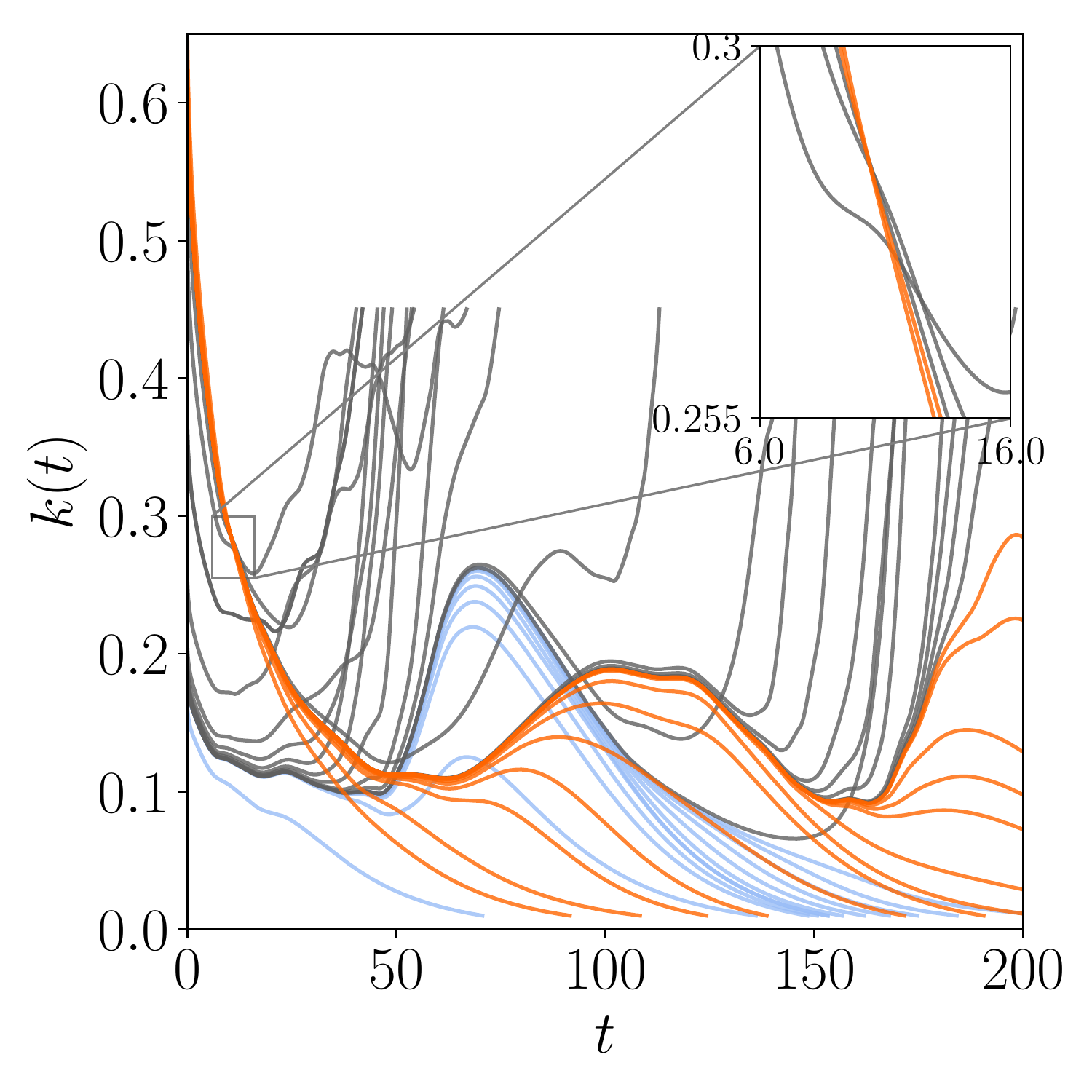}
	(c)
	\begin{picture}(1,1.05373823)%
	\put(0,0){\includegraphics[width=\unitlength,page=1]{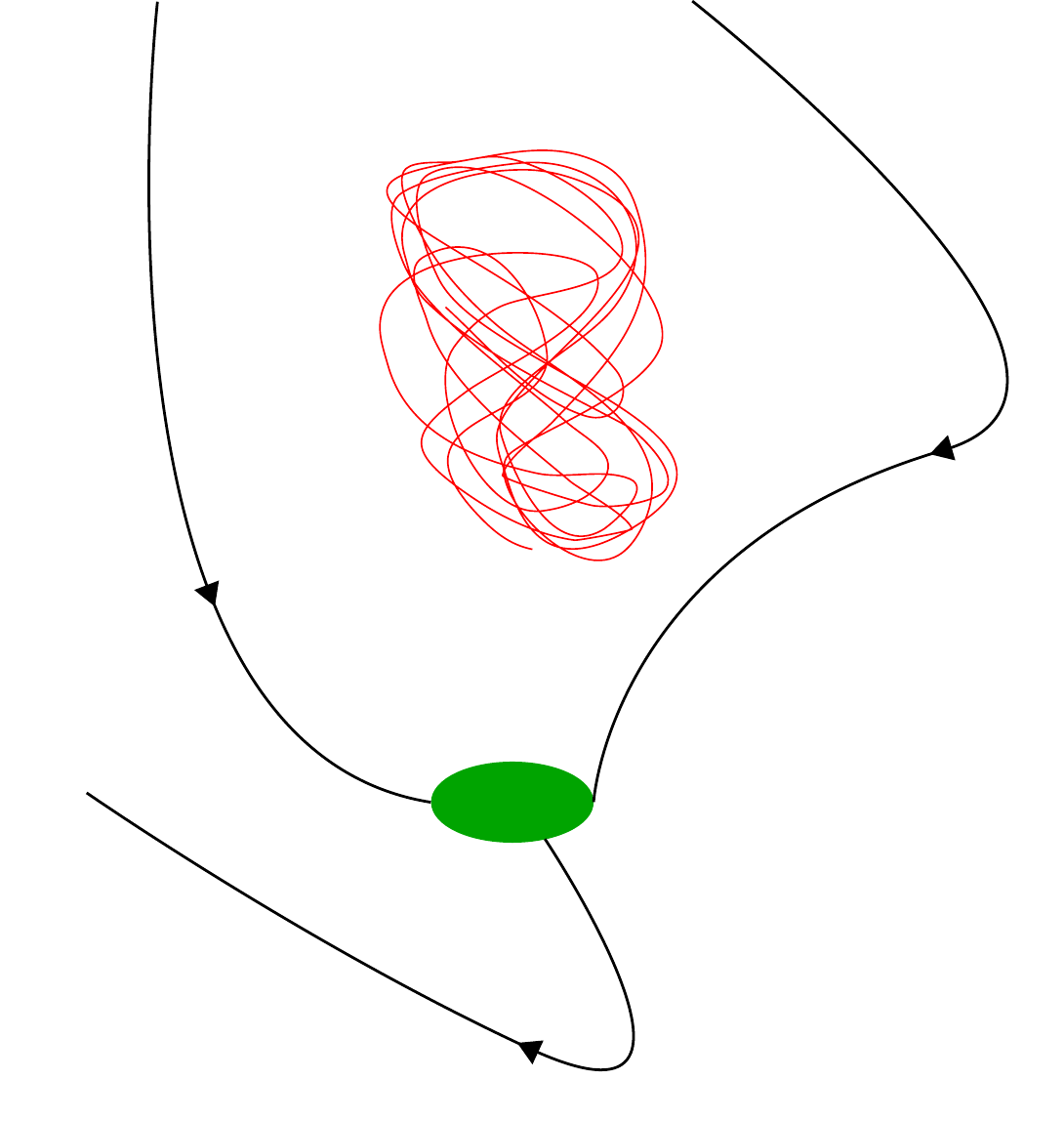}}%
	\put(0.16303317,0.79689392){\color[rgb]{0,0,0}\makebox(0,0)[lt]{\begin{minipage}{0.07651899\unitlength}\raggedright \end{minipage}}}%
	\put(0,0){\includegraphics[width=\unitlength,page=2]{edgesketchnocont}}%
	\put(0.37265112,0.96825234){\color[rgb]{0,0,0}\makebox(0,0)[lt]{\begin{minipage}{0.38259494\unitlength}\raggedright Turbulence\end{minipage}}}%
	\put(0,0){\includegraphics[width=\unitlength,page=3]{edgesketchnocont}}%
	\put(-0.00375496,0.38990222){\color[rgb]{0,0,0}\makebox(0,0)[lt]{\begin{minipage}{0.40810127\unitlength}\raggedright Laminar\end{minipage}}}%
	\put(0.10274631,0.54388144){\color[rgb]{0,0,0}\makebox(0,0)[lt]{\begin{minipage}{0.22955696\unitlength}\raggedright $W_s$\end{minipage}}}%
	\put(0,0){\includegraphics[width=\unitlength,page=4]{edgesketchnocont}}%
	\put(0.915,0.82){\color[rgb]{0,0,0}\makebox(0,0)[lt]{\begin{minipage}{0.03825949\unitlength}\raggedright $A$\end{minipage}}}%
	\put(0.23283958,0.46256875){\color[rgb]{0,0,0}\makebox(0,0)[lt]{\begin{minipage}{0.03825949\unitlength}\raggedright $B$\end{minipage}}}%
	\put(0,0){\includegraphics[width=\unitlength,page=5]{edgesketchnocont}}%
	\put(0.50864098,0.11255198){\color[rgb]{0,0,0}\makebox(0,0)[lt]{\begin{minipage}{0.10202532\unitlength}\raggedright $W_u$\end{minipage}}}%
	\put(0.45,0.32){\color[rgb]{0,0,0}\makebox(0,0)[lt]{\begin{minipage}{0.40810127\unitlength}\raggedright \tiny{Edge} \end{minipage}}}%
	\put(0.45,0.293){\color[rgb]{0,0,0}\makebox(0,0)[lt]{\begin{minipage}{0.40810127\unitlength}\raggedright \tiny{state} \end{minipage}}}%
	\put(0,0){\includegraphics[width=\unitlength,page=6]{edgesketchnocont}}%
	\put(0.32552984,0.63382821){\color[rgb]{0,0,0}\makebox(0,0)[lt]{\begin{minipage}{0.10202532\unitlength}\raggedright $W_u$\end{minipage}}}%
	\put(0,0){\includegraphics[width=\unitlength,page=7]{edgesketchnocont}}%
	\put(0.83,0.61){\color[rgb]{0,0,0}\makebox(0,0)[lt]{\begin{minipage}{0.22955696\unitlength}\raggedright $W_s$\end{minipage}}}%
	\end{picture}%
	\caption{
		(a) Time-series of perturbation kinetic energy for five initial 
		conditions from $\Reynolds = 10000$. 
		Annotated snapshots are the flow structures 
		for one of the simulations at times $t = 0.0, 25, 50$. 
		Visualized here are the isosurfaces of streamwise velocity at $50\%$
		of its maxima and minima (red and blue) as well as the streamwise
		vorticity isosurfaces at $25\%$ of its maxima and minima (green and 
		purple) at the respective instances.         
		(b) Time evolution of perturbation kinetic energy for initial 
		conditions obtained from rescaling turbulent initial conditions at $\Reynolds=10000$ 
		such that trajectories neither becomes turbulent nor laminarize 
		for longer and longer times.
		Inset: Zoom-in to the region $\zeit \in [0, 16]$ and 
		$k \in [0.255, 0.3]$.
		(c) A ``cartoon'' of the \statesp\ where initial conditions that 
		relaminarize are separated from those that develop into turbulence by the 
		stable manifold of the ``edge state''.
		\label{f-edgeUp}}
\end{figure}

We utilize the \texttt{Openpipeflow} solver\rfp{willis-2017}
to simulate pipe flow in an axially-periodic computational domain of 
length $L = 2 \pi / 0.625 \approx 10R$, where $R$ is the pipe radius. 
The Reynolds number is set to 
$\Reynolds = U_c R / \nu = 3000$, where $U_c$ is the centerline 
velocity, $\nu$ is the kinematic viscosity, and the resolution is 
identical to the one used in Budanur \& Hof\rfp{budanur-hof-2018}. 
In results, the time scale is chosen to be advective unit 
$4 R / U_c$.
In numerical simulations, a convenient method for obtaining initial 
conditions that have flatter mean profiles than normal levels at a 
certain \Reynolds\ is to take typical turbulent states from higher 
\Reynolds . 
\refFig{f-edgeUp} (a) shows the time-evolution of the 
perturbation kinetic energy of the velocity fluctuations 
with respect to the laminar flow,
from five simulations at $\Reynolds = 3000$ where 
we used typical turbulent states from $\Reynolds = 10000$ as 
initial conditions.
The flow structures visualized in 
\reffig{f-edgeUp} correspond to three snapshots from one of the 
simulations and they illustrate the laminarization of the 
highly-turbulent initial state. We carried out edge-tracking 
starting with one of these states at $\Reynolds=10000$. Let us 
denote it by $\ssp_0$.
\refFig{f-edgeUp} (b) shows the time-evolution of the 
kinetic energy when initial conditions $\ssp(0) = c \ssp_{0}$, 
with $c \in (0, 1]$, evolve at $\Reynolds = 3000$. 
The 
time-series data on \reffig{f-edgeUp} (b) was produced as follows: 
we first set $c = 0.5, 0.75, 1.0$ and forward-integrated these initial
conditions and observed that $c = 0.5$ and $c = 1.0$ 
laminarized while the initial condition with $c = 0.75$ became
turbulent. We then carried out two bisection searches, akin to that of
\refref{IT01}, for $c \in (0.5, 0.75)$ and $c \in (0.75, 1.0)$ such that
simulations neither become turbulent nor laminarize. On 
\reffig{f-edgeUp} (b), the time series data with initial conditions 
$c \geq 0.75$ are shown orange while those with $c < 0.75$ are 
plotted blue. As shown in the time-series plots, there appear to be two 
boundaries between the laminarizing trajectories and the ones that 
trigger turbulence. The low-energy boundary (blue) separates the initial
conditions that are strong enough to trigger turbulence from those 
that are too weak. In contrast, the high-energy boundary (orange) 
separates the initial conditions that are too strong to trigger 
turbulence from those that are weak enough. We will refer this latter
boundary as \textit{the upper edge of chaos}. Both trajectories obtained
from the bisection end up in the vicinity of the same edge state, 
characteristic of which are described by \refref{budanur-hof-2018}. 
\notetarget{fig1c}{In} \reffig{f-edgeUp} (c),
 we show a state-space sketch illustrating 
this numerical experiment, where 
     the Euclidean distance from the laminar state should be interpreted as 
	 the square root of the perturbation energy and the stable and 
 	 unstable manifolds of the edge state are annotated with 
   	 $W_s$ and $W_u$, respectively. 
   	 The initial conditions of our numerical
   	 experiment lie along the dashed line in \reffig{f-edgeUp} (c),
   	 and the high-energy intersection $A$ of this line with the stable 
   	 manifold $W_s$ of the edge state marks the upper edge of chaos. 

\begin{figure}
	\centering
	(a) \includegraphics[width=0.45\textwidth]{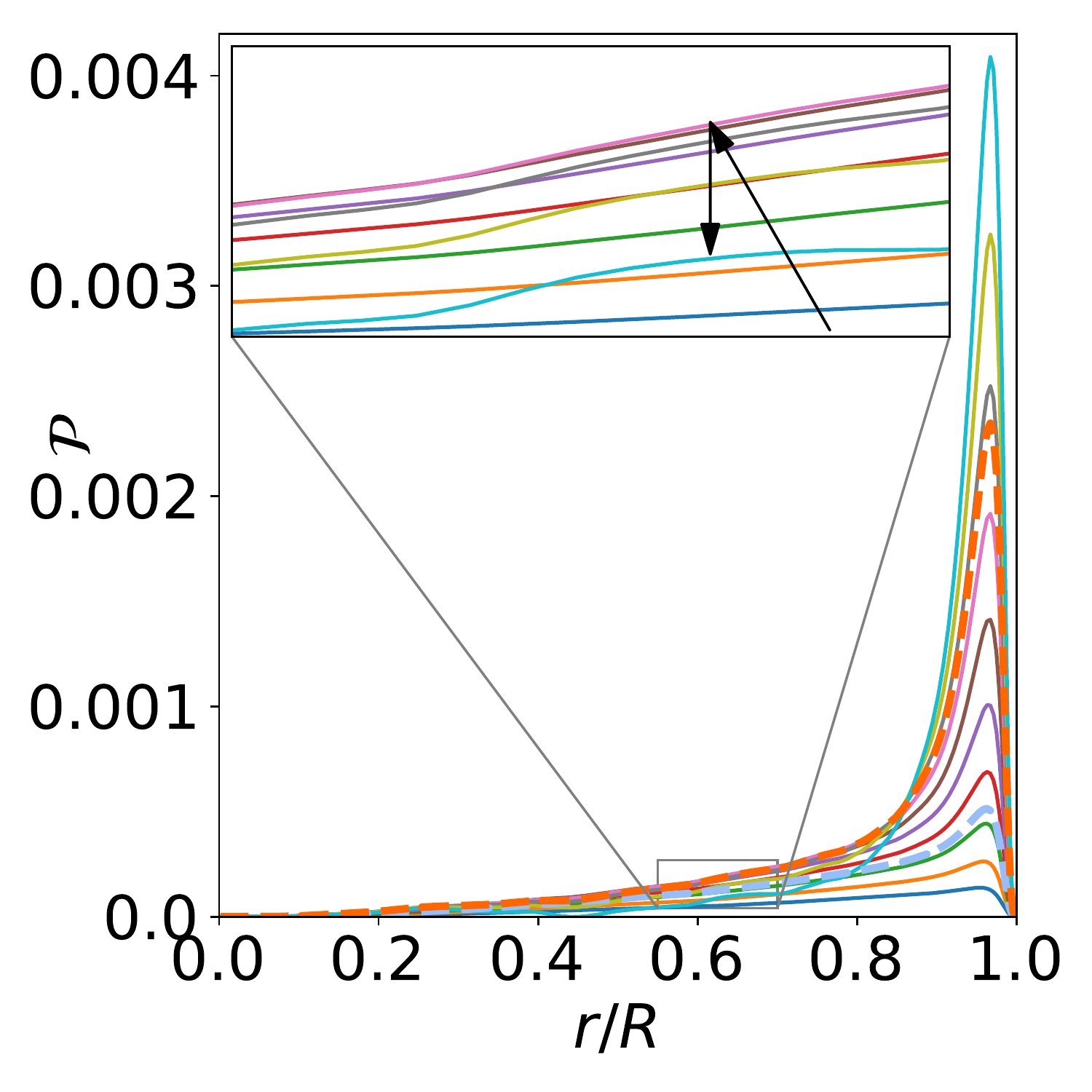}
	(b) \includegraphics[width=0.45\textwidth]{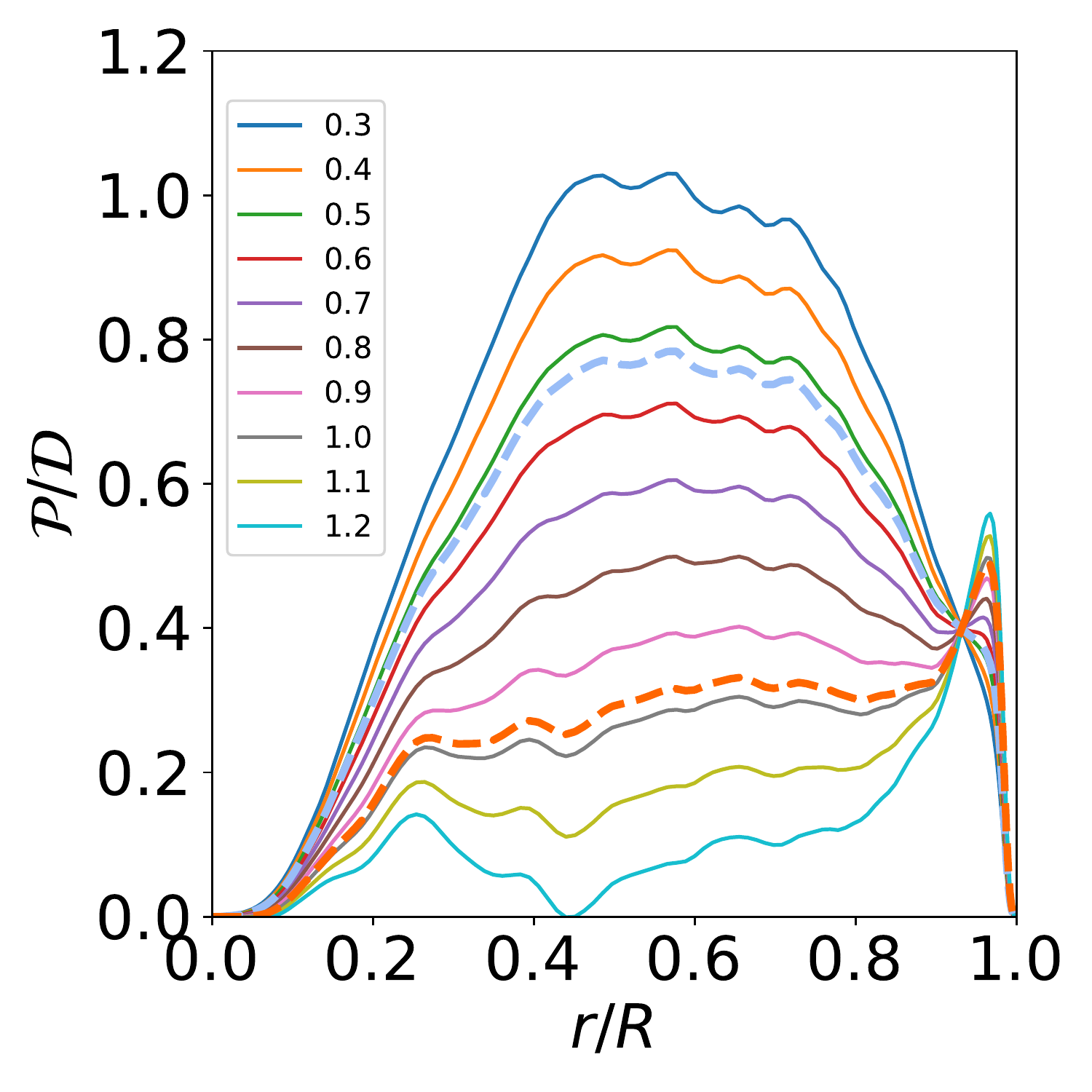} \\ 
	(c) \includegraphics[width=0.45\textwidth]{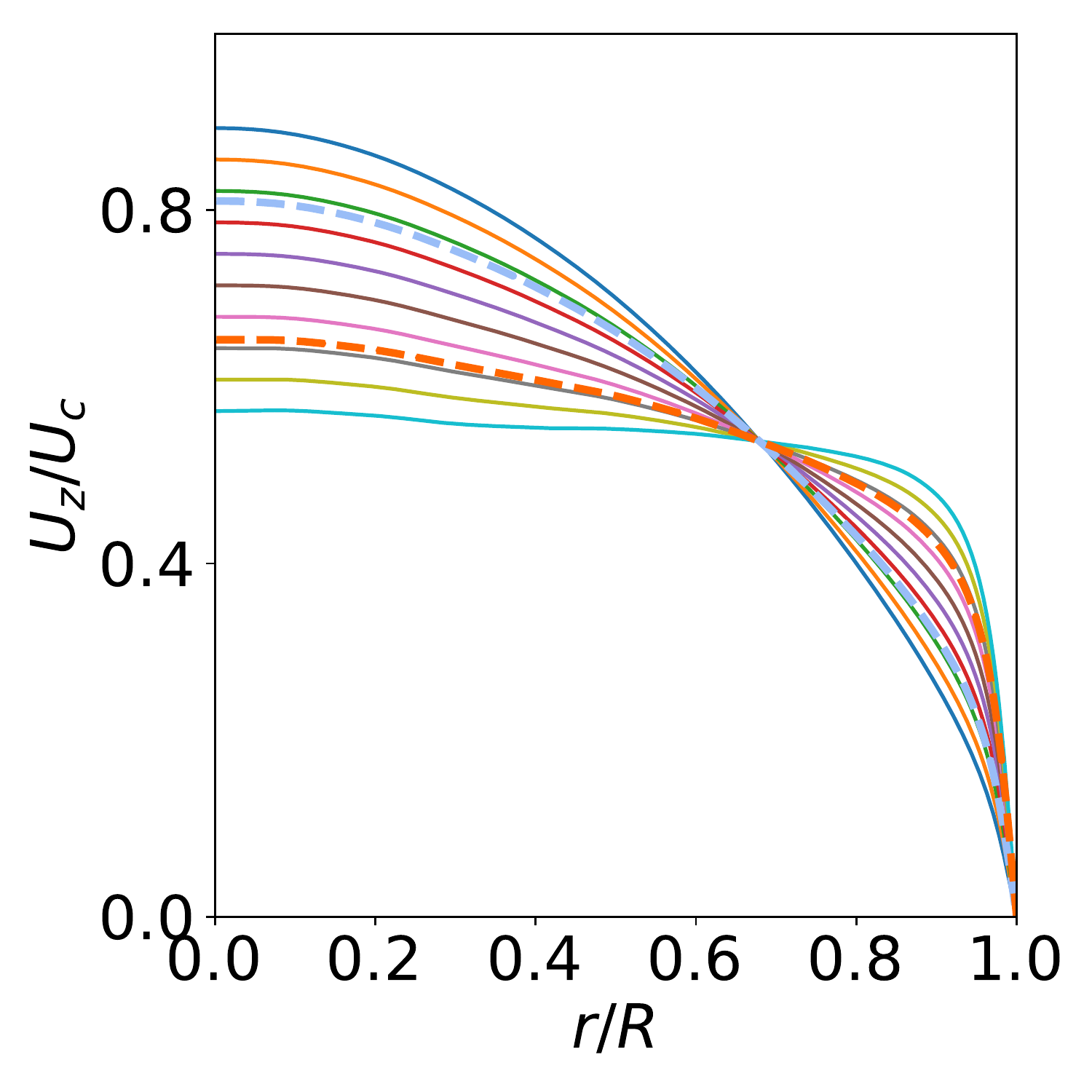}
	(d) \includegraphics[width=0.45\textwidth]{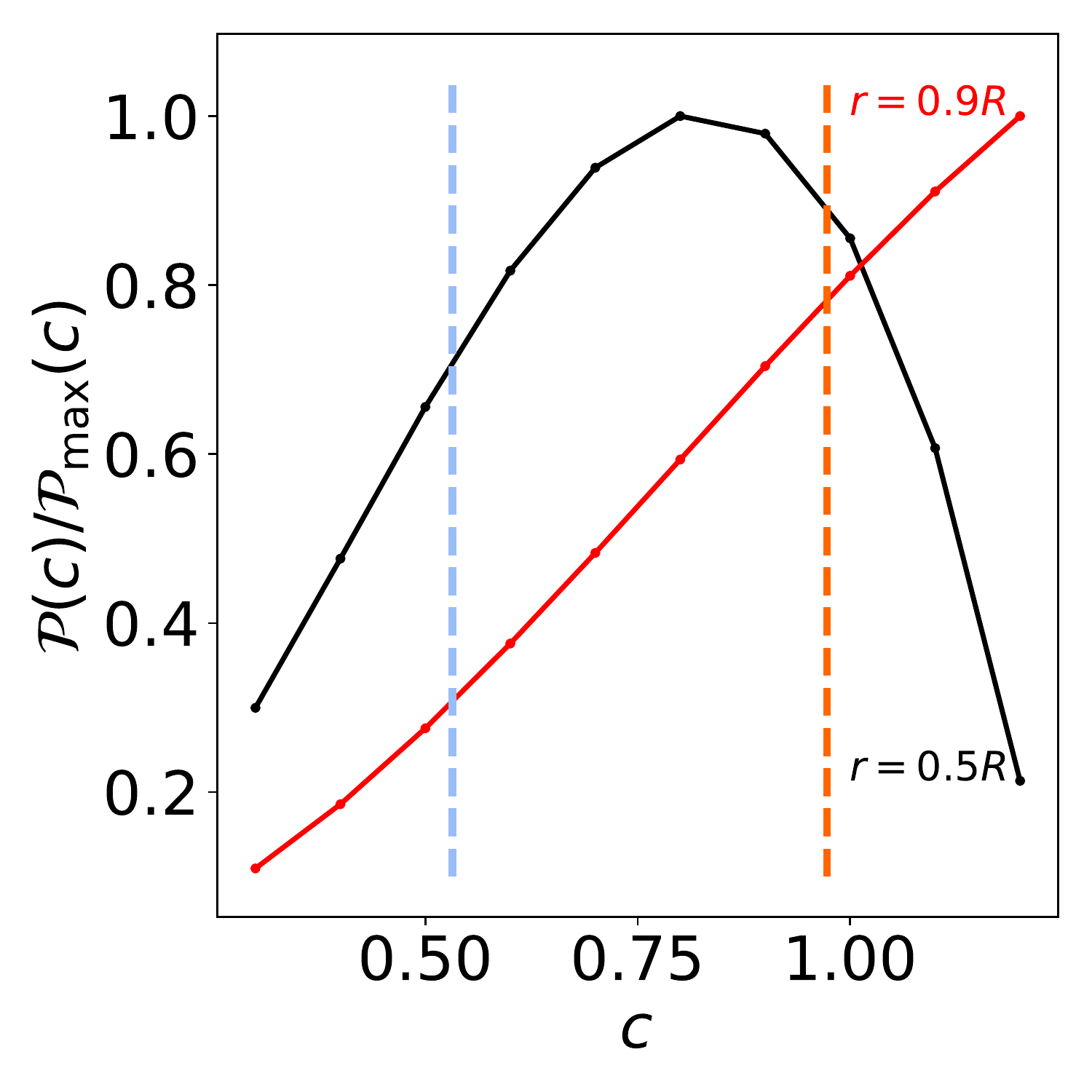}
	\caption{(a) $\product$, 
		(b) $\product/\dissip$, 
		(c) $U_z$, 
		as a function of 
		radial position for initial conditions $\ssp(0)=c \ssp_{0}$, 
        where $\ssp_{0}$ is a turbulent state at $\Reynolds = 10000$ 
        and $c \in [0.3, 1.2]$.
		Different colors correspond to different $c$ values shown in 
        the legend of the panel (b). 
        Blue and orange dashed curves correspond to the critical 
        initial conditions respectively on the lower and the upper edge 
		of chaos with corresponding coefficients $c = 0.531856455058$ 
        and $c = 0.972761377146$.
		The arrows in the inset of (a) are there to emphasize that the 
        production in the bulk region decreases for $c>0.9$.
		(d) $\product$ as a function of $c$ at $r = 0.5R$ and 
        $r = 0.9R$, normalized by the respective maxima of the curves. 
        Dots correspond to the data points and they are connected with 
        line segments in order to guide the eye.
		\label{f-prodisp}}  
\end{figure}

For high-energy turbulent initial conditions to laminarize, they 
must experience substantial energy losses. We are now going to 
examine where these losses take place by measuring turbulent
energy production and dissipation as a function of the 
radial coordinate. 
When the turbulent velocity field is 
decomposed into its mean $\langle \bm{U} \rangle$ and fluctuating parts
$\vfield$, rates of the turbulent energy production $\product$ 
and the dissipation $\dissip$ can be expressed as follows 
\rfp{popebook}
\beq
\mathcal{P} = - \langle u_i u_j \rangle 
\frac{\partial \langle U_i \rangle}{\partial x_j}
\, , 
\quad 
\dissip = 2 \nu \langle s_{ij} s_{ij} \rangle \, ,
\label{eq:ProDisp}
\eeq
where $s_{ij} = (\partial_i u_j + \partial_j u_i) / 2$ is the fluctuating
rates of strain tensor and $\partial_i = \partial / \partial x_i$. 
Hence, turbulent dissipation can also be expressed as 
\beq
\dissip = \nu \langle \partial_i u_j \partial_i u_j +
\partial_i u_j \partial_j u_i \rangle \, . 
\label{eq:Disp}
\eeq
It can be readily seen from \refeq{eq:Disp} that scaling up the 
amplitude of velocity fluctuations increases the dissipation
everywhere. The production, on the other hand, has a different 
trend: as we have visualized in \reffig{f-prodisp} (a), 
scaling up the velocity fluctuations 
increases the production near the pipe wall ($r \approx R$), 
however, away from the wall, this trend reverses. 
As a result, the bulk region of the pipe becomes completely dissipative
as illustrated by
the radial profiles of  $\product/\dissip$, shown in \reffig{f-prodisp} \notetarget{typo1}{(b)}.
This is due to the gradient term in 
the production equation \refeq{eq:ProDisp} and flattening of the
velocity profile for highly turbulent initial conditions as shown
in \reffig{f-prodisp} (c). Figure \ref{f-prodisp}(d) shows the normalized production 
at $r = 0.5R$ and $r = 0.9R$ as a function of the scaling coefficient $c$. 
While the production near the wall increases monotonically
with $c$, the production in the bulk peaks in the range of 
$c$ that lead to transition,
delimited by the vertical dashed blue (lower edge) and orange (upper edge) lines. 
For values of $c$ that lead to transition, the maximum of the normalized production is 
in the bulk region of the pipe, 
thus \notetarget{fig2disc}{suggesting}
the importance of fluctuations away from the pipe wall for transition. 

Open-loop control strategies that aim to laminarize turbulent
	flow, such as the ones employed by 
	K\"uhnen \etal \rf{kuhnen-etal-2018a}, must 
	achieve two goals: (1) manipulate the flow in such a way 
	that its structures are dissipative at a given \Reynolds , 
	(2) avoid retransition to turbulence.
	We hypothesize that the flattening of mean profile and 
	resulting decrease of production in the bulk region
	ensures the second condition by eliminating all 
	fluctuations in the bulk.
As a first test of this hypothesis,
we removed the mean of the axial velocity 
	 perturbations from the initial conditions on the laminarizing 
	 side of the upper edge, 
	 and rerun our numerical experiments. 
	 Albeit being energetically weaker, these
	 initial conditions had velocity profiles equal to that of the 
	 laminar flow and they triggered the transition after an initial 
	 small energy drop.
As a further test,
we performed simulations with random solenoidal 
initial conditions that do not alter the mean profile. To this end,
we populated a subset 
of modes in the spectral expansion of the velocity field 
(excluding the mean-flow distortion)
with random amplitudes and ran the pressure solver on this state in order
to ensure that it is divergence-free; see \refref{marensi-etal-2019} for 
details. Given they are strong enough, all initial conditions 
constructed this way lead to a transition to turbulence and we did 
not observe an upper edge of chaos in this case.

Turbulent wall-bounded shear flows are in equilibrium in a statistical 
sense. That is, on average, an equal amount of turbulent kinetic energy is 
produced and dissipated. It is well known \rfp{popebook, KMM87, YWH2018} 
that the majority of the production takes place near the wall, 
where the shear is largest, and the flow becomes strongly dissipative
away from the wall. The edge state of the pipe flow under consideration
accommodates traveling wave solutions, which are not only 
statistically sustained
but also dynamically invariant. In other words, they are in 
energy balance at all times. In the next section, we
demonstrate that the energy budget of these 
lower-branch traveling waves or edge
solutions is remarkably 
different from that of the turbulence, and the production away from 
the wall plays an essential role in sustaining them.

\section{Energetics of the edge state}

\begin{figure}
	\centering
	(a) \includegraphics[width=0.29\textwidth]{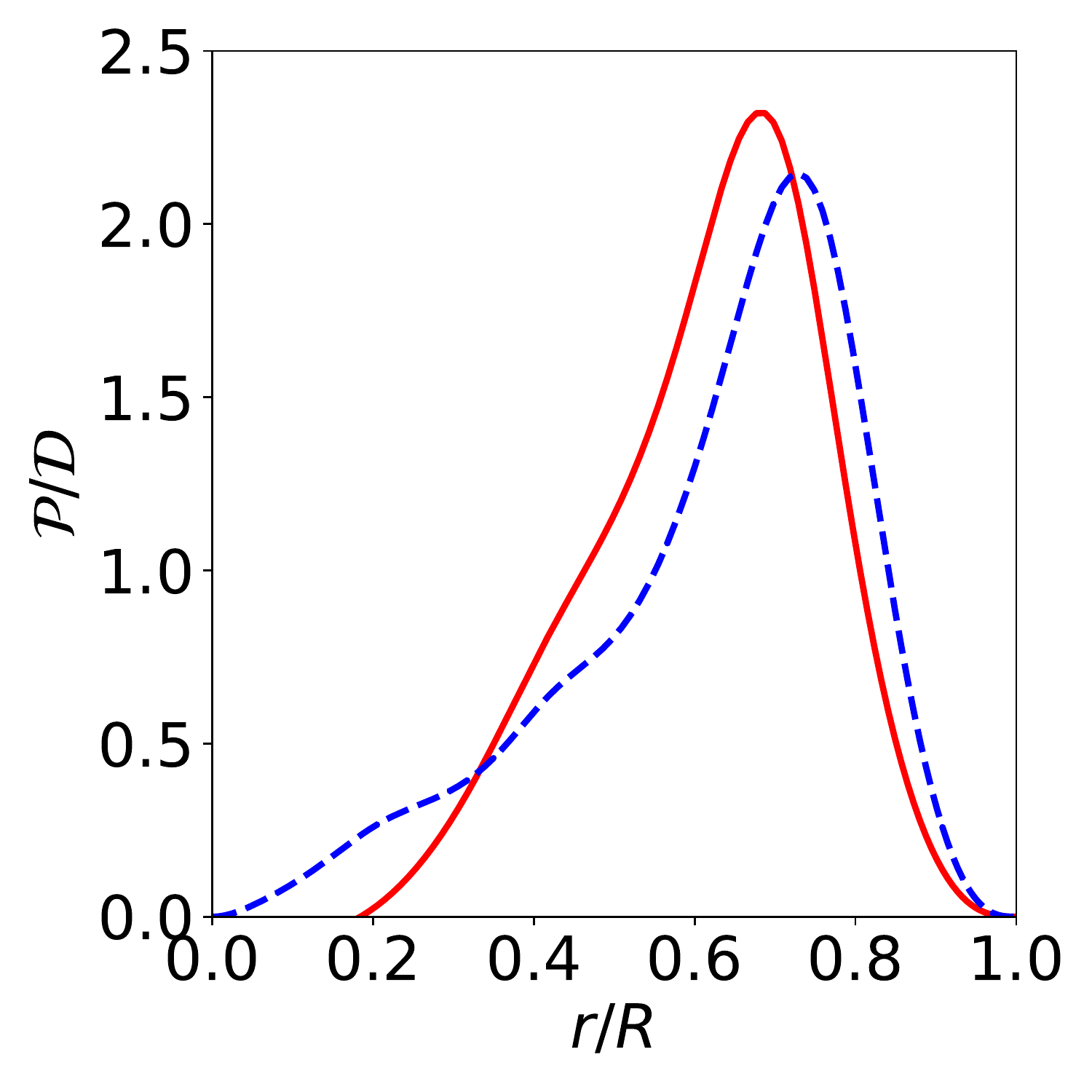}
	(b) \includegraphics[width=0.29\textwidth]{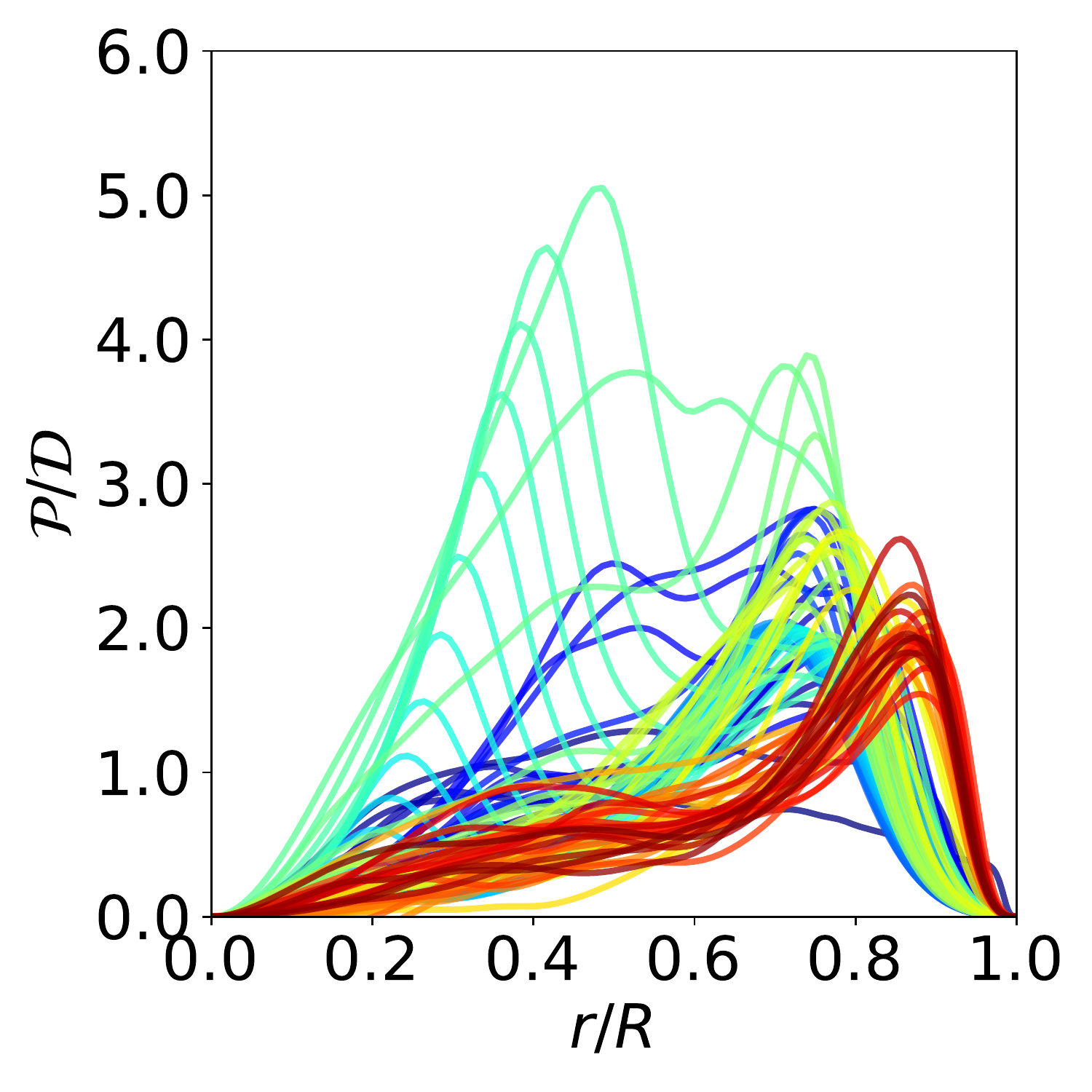}
	(c) \includegraphics[width=0.29\textwidth]{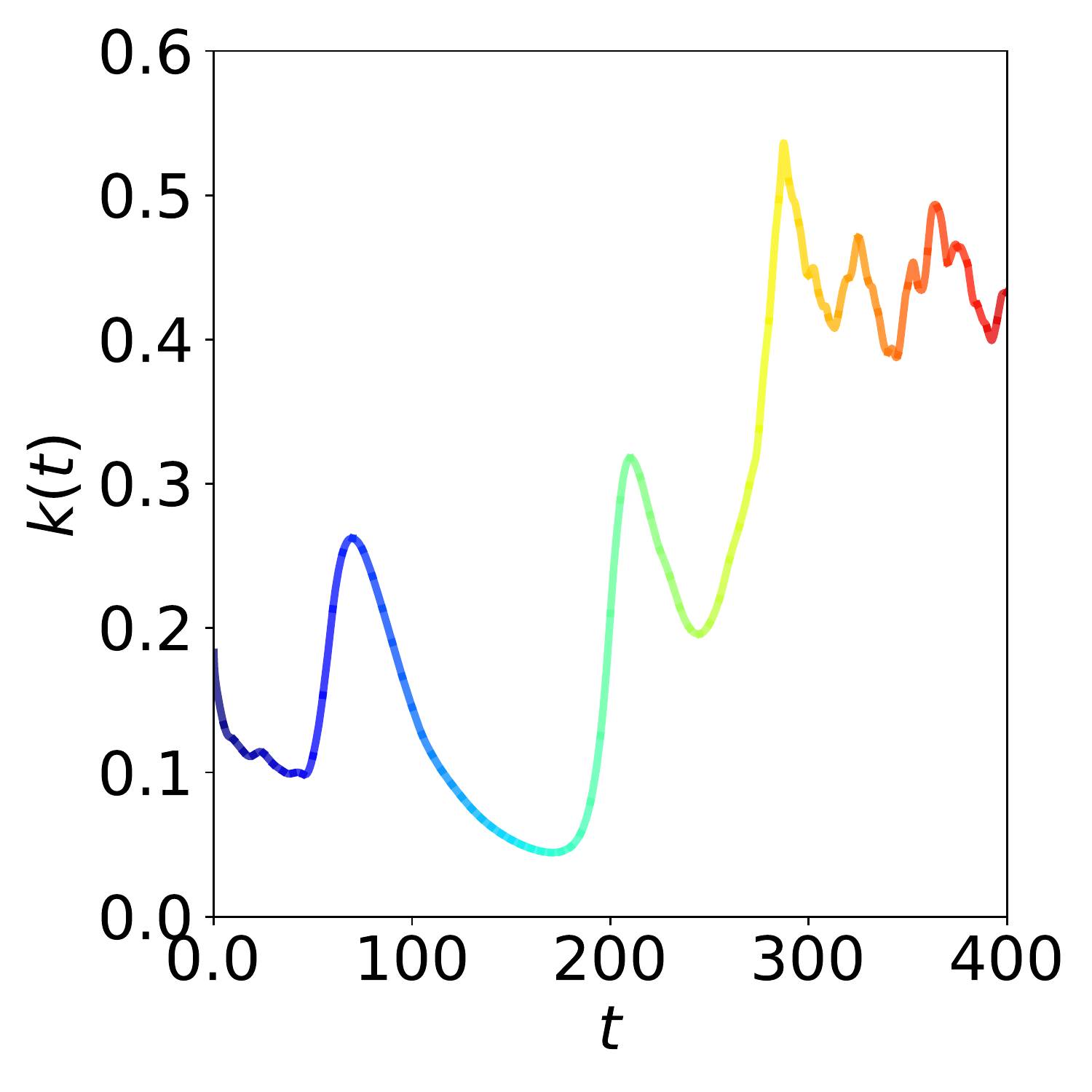} 
	\caption{
		(a) $[\product/\dissip] (r)$ for traveling waves $S_1$ 
		(red/solid) and 
		$S_{1N}$ (blue/dashed) on the laminar-turbulent boundary.
		(b) Evolution of $[\product/\dissip] (r)$ as a trajectory proceeds 
		towards turbulence after spending a long time on the ``edge''. 
		Colors indicate the direction of time (initial:blue, final:red); 
		each curve is separated by $\Delta t = 5 $ in time. 
		(c) Time-series of perturbation kinetic energy with the 
		same color-coding for 
		reference.
		\label{f-prodispt}}  
\end{figure}

The edge state of a shear flow is considered to be dynamically 
invariant, that is, if an initial condition starts exactly on the 
edge state, it should stay there forever. When the edge state is 
stationary\rfp{IT01}, or time-periodic\rfp{TI03,AvMeRoHo13} this 
description can be supported by computing numerically exact 
invariant solution by a Newton method. When the edge state is 
chaotic as it is the case here, 
such a computation is not possible; and thus, 
the edge state can only be followed by edge tracking for finite
times, as shown in \reffig{f-edgeUp} (b). Even though the stable and
unstable manifolds of chaotic sets are not well defined, the 
edge-tracking observations that are reported in the previous 
literature\rfp{SchEckYor07,duguet07,budanur-hof-2018} and 
the current work agree with the hypothesis that the edge state
of the pipe flow is an invariant set of saddle-type. Moreover, 
\refrefs{duguet07} and\rf{budanur-hof-2018} reported that edge 
tracking trajectories approach traveling waves 
of the pipe flow, which are by definition time-invariant.
\refFig{f-prodispt}(a) shows the $[\product/\dissip]$ ratio as 
a function of radius for the 
traveling waves $S_1$ and $S_{1N}$, 
which are embedded in the laminar-turbulent boundary\rfp{budanur-hof-2018}. 
For the traveling waves $S_1$ and $S_{1N}$, 
the $[\product / \dissip](r)$ 
curve peaks
at $r = 0.69R$ and $r = 0.73R$ respectively, which, 
is quite different from the turbulent case.

In order to see how generic trajectories on the laminar turbulent boundary 
are sustained, we measured $[\product/\dissip](r)$ along an 
edge-tracking trajectory, which proceeds towards turbulence after spending 
a long time on the laminar turbulent boundary. 
\refFig{f-prodispt}(b) shows the time-evolution of 
$[\product/\dissip](r)$ measured at instances separated by 
$\Delta t = 5$ and the time is color-coded as in the 
time-series of \reffig{f-prodispt} (c). Once the turbulence is 
statistically stationary ($t>300$, red curves on 
\reffig{f-prodispt}(b)), the $[\product/\dissip](r)$ peak settles in the
near-wall region ($r > 0.8R$). Prior to this, when the trajectory is 
exploring the edge-state, $[\product/\dissip](r)$ has a completely 
different radial-profile with one or more maxima
away from the wall. Similar to the traveling waves on the 
laminar-turbulent boundary, the chaotic edge state is also sustained 
by the production away from the wall.

\section{Transition from the minimal seed}

\begin{figure}
  \centering
    (a) \includegraphics[width=0.45\textwidth]{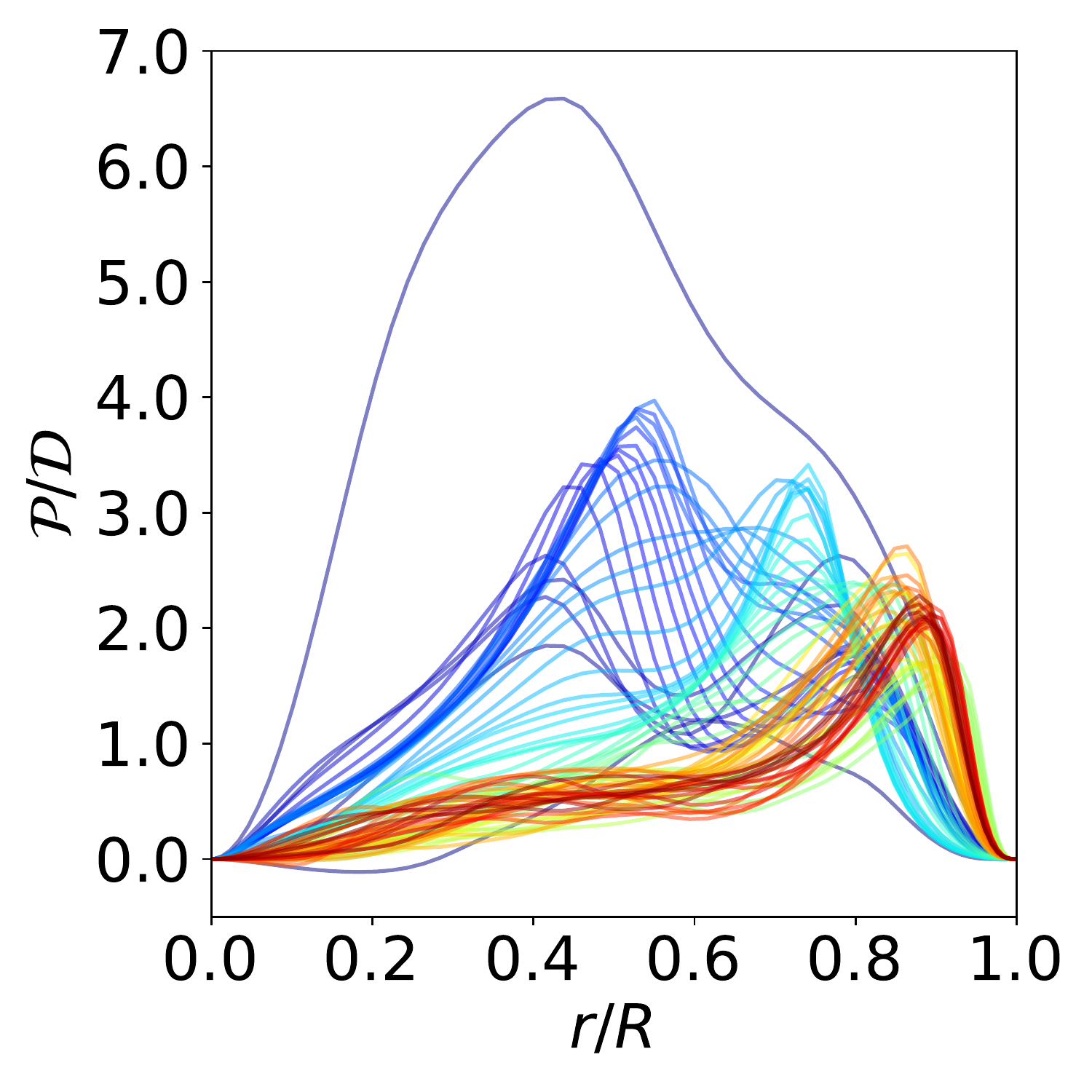}
	(b) \includegraphics[width=0.45\textwidth]{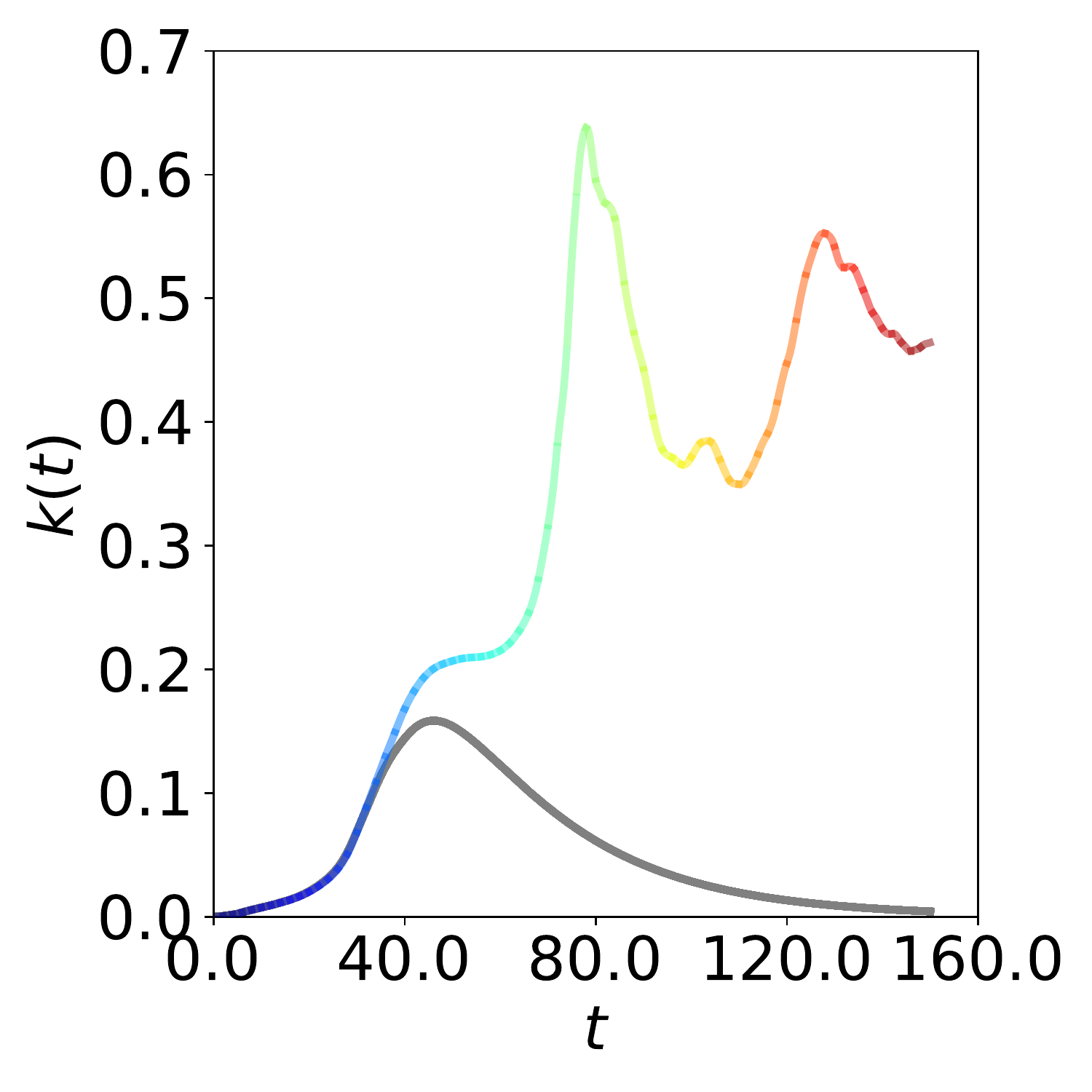} 
	  \caption{
      (a) Time evolution of $[\product/\dissip] (r)$ for the minimal seed
	 at $Re=3000$.
      Colors indicate the direction of time (initial:blue, final:red); 
      each curve is separated by $\Delta t = 2$ in time. 
      (b) Time-series of perturbation
      kinetic energy with the same color-coding for 
      reference. The gray line corresponds to the energy time series for the largest tested value of $E_0$ below which transition never occurred. 
       \label{fig-PDtime-MS}}  
\end{figure}
Refs.\rf{pringle-kerswell-2010,pringle-etal-2012,marensi-etal-2019} 
provided evidence that the nonlinear optimal in pipe flow tracks the 
laminar-turbulent boundary before either relaminarizing 
or triggering turbulence.
We analyze the $[\product/\dissip](r)$ curves at different stages of 
the transition from the minimal seed in order to reveal the radial 
locations of the strong energy amplification.

We calculate the minimal seed at $Re=3000$, with two-digits accuracy in 
the critical initial energy. 
Following previous studies (e.g.\rfp{pringle-etal-2012}), an 
optimization horizon $T_{opt}=75$ is chosen.
\refFig{fig-PDtime-MS}(a)
shows a family of curves 
$[\product/\dissip](r)$ along the minimal seed trajectory, separated by 
time intervals of
$\Delta t = 2$ and colored according to the coding given in 
\reffig{fig-PDtime-MS}(b).
The latter graph shows the 
time-series of the kinetic energy for the two initial conditions that 
bracket the minimal seed for transition, up to the chosen accuracy. 
Analogous to \reffig{f-prodispt}(b), once the turbulent attractor is 
reached (red curves in \reffig{fig-PDtime-MS}(a)) the peak of 
$[\product/\dissip](r)$ settles in the region close to the wall, while 
in the initial growth phase, up to the edge and when tracking the edge 
(blue curves in \reffig{fig-PDtime-MS}(a)), the peak of 
$[\product/\dissip](r)$ is closer to the pipe center.
Similar to the traveling waves embedded in the laminar-turbulent 
boundary and to the chaotic edge, the minimal seed is also strongly 
amplified in the bulk region.
\refFig{fig-PDtime-MS} thus supports our picture that the (large-scale) 
structures away from the the pipe wall are more important than the 
(small-scale) structures close to the wall for the transition process, 
while, after turbulence is triggered, the near-wall structures become 
dominant.

Particularly interesting is the significant increase of 
$\product/\dissip$ experienced by the minimal seed during its initial 
evolution, up to the edge; e.g. refer to the blue curve almost reaching 
$\product/\dissip \approx 7$ in \reffig{fig-PDtime-MS}. To better 
understand the initial mechanism of growth of the minimal seed, we 
closely analyze the $[\product/\dissip](r)$ profiles in the time window 
$0 \le t \le 5$. \refFigs{fig-PDtime-MS_t5}(a, b) show a zoom of 
\reffigs{fig-PDtime-MS}(a, b) in this time window, with intervals of 
$\Delta t = 0.1$. The time evolution of the peak 
$(\product/\dissip)_{max}$ of the production to dissipation ratio and 
of its radial position $r_{max}$ are shown in 
figure \ref{fig-PDtime-MS_t5}(c). 
As outlined in \S \ref{sec:intro}, in the Orr and oblique-wave phases 
of its initial evolution, the minimal seed gradually delocalizes and 
undergoes most of its energy growth. This is clearly seen in 
figure \ref{fig-PDtime-MS_t5}(c) by the marked increase 
of $(\product/\dissip)_{max}$ and by the shift of $r_{max}$ towards the 
pipe center. The peak $(\product/\dissip)_{max} \approx 12$ is reached 
at $t\approx 1.5$ and occurs almost at $r\approx0.4R$.
After this initial spurt of energy and delocalization, 
$(\product/\dissip)_{max}$ approaches the typical `turbulent' value of 
$\approx 1.8$ \rfp{popebook} and moves closer to the wall, i.e. 
$r_{max}\approx0.8R$.
\citet{marensi-etal-2019} showed that the minimal seed at the end of 
the Orr and oblique-wave phases of its evolution is a useful tool to 
measure transition thresholds for typical ambient disturbances. 
The Orr process ($t \le 1-1.5$ for $Re =2400 -3500$) was identified by 
analyzing the flow topology of the minimal seed at the very initial 
stage of its evolution, but, as pointed out by the authors themselves, 
there was some discretion in the identification of this phase. The 
oblique-wave phase ($t \le 2.5-3$ for the same range of $Re$) is 
usually signaled by a `bump' in the time evolution of the three 
dimensional energy $E_{3d}$, i.e. the energy associated with the 
streamwise dependent modes only. However, in a long pipe this `bump' is 
obscured by the long wave-length modes, which thus need to be filtered 
out first.
Here, while analyzing the energetics of transition in pipe flow, we 
realized that the Orr and oblique-wave phases of the transition from 
the minimal seed could be identified in a much easier and clearer way 
using the curves of $[\product/\dissip](r)$.
In particular, a graph like that shown in figure 
\ref{fig-PDtime-MS_t5}(c), indicates that the Orr phase ends at 
$t \approx 1$ where $(\product/\dissip)_{max}$ reaches its peak, while 
the end of the oblique-wave phase is signaled by the jump of $r_{max}$ 
from close to the pipe center to close to the wall (and, 
correspondingly, the marked change of slope of 
$(\product/\dissip)_{max}$) which occurs at $t \approx 2.5-3 $.
The analysis of the $[\product/\dissip](r)$ profiles is thus found to 
provide a simple and clear method for identifying the Orr and 
oblique-wave phases of the minimal seed evolution. 
\section{Implications for flow control}

\begin{figure}
  \centering
    (a) \includegraphics[width=0.29\textwidth]{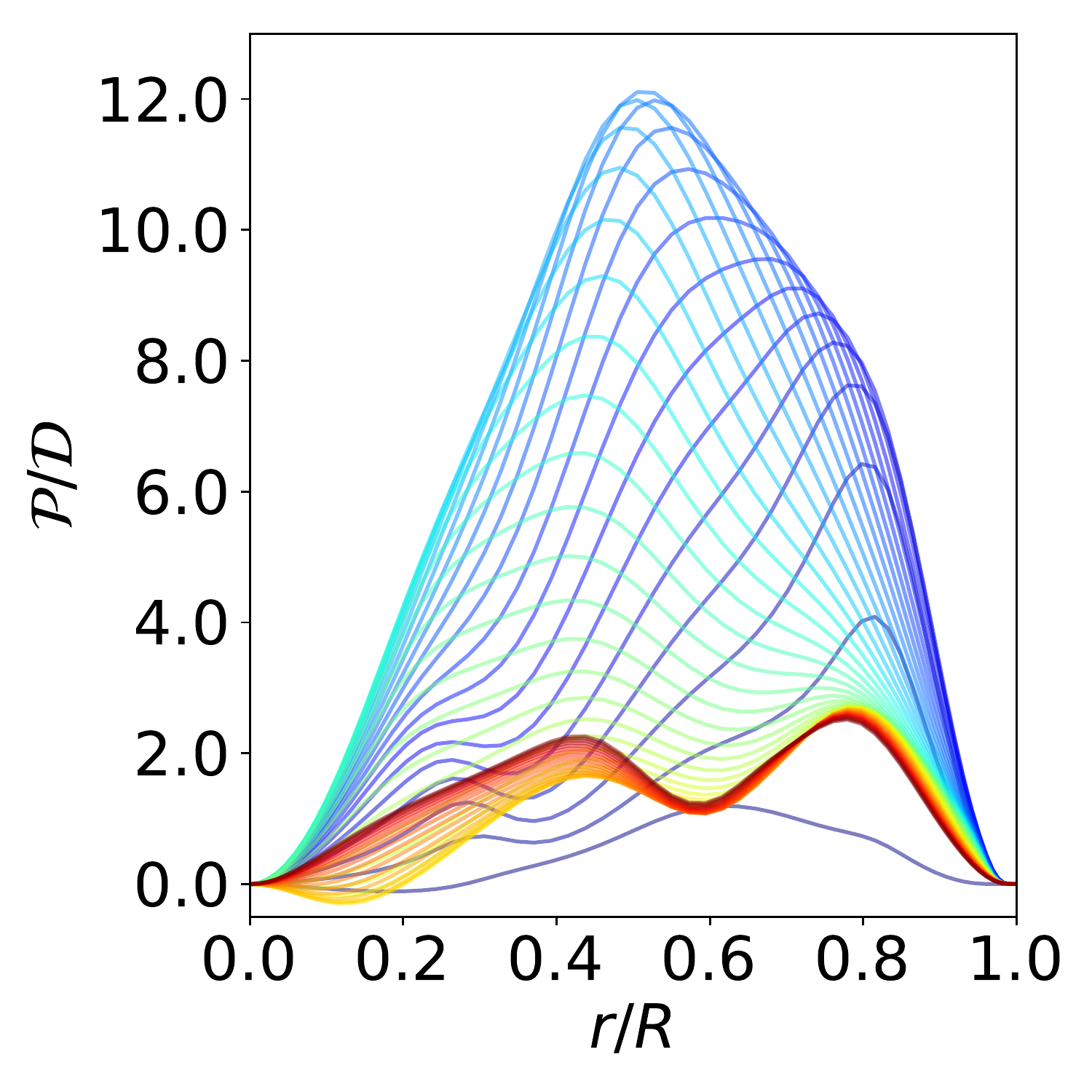}
	(b) \includegraphics[width=0.29\textwidth]{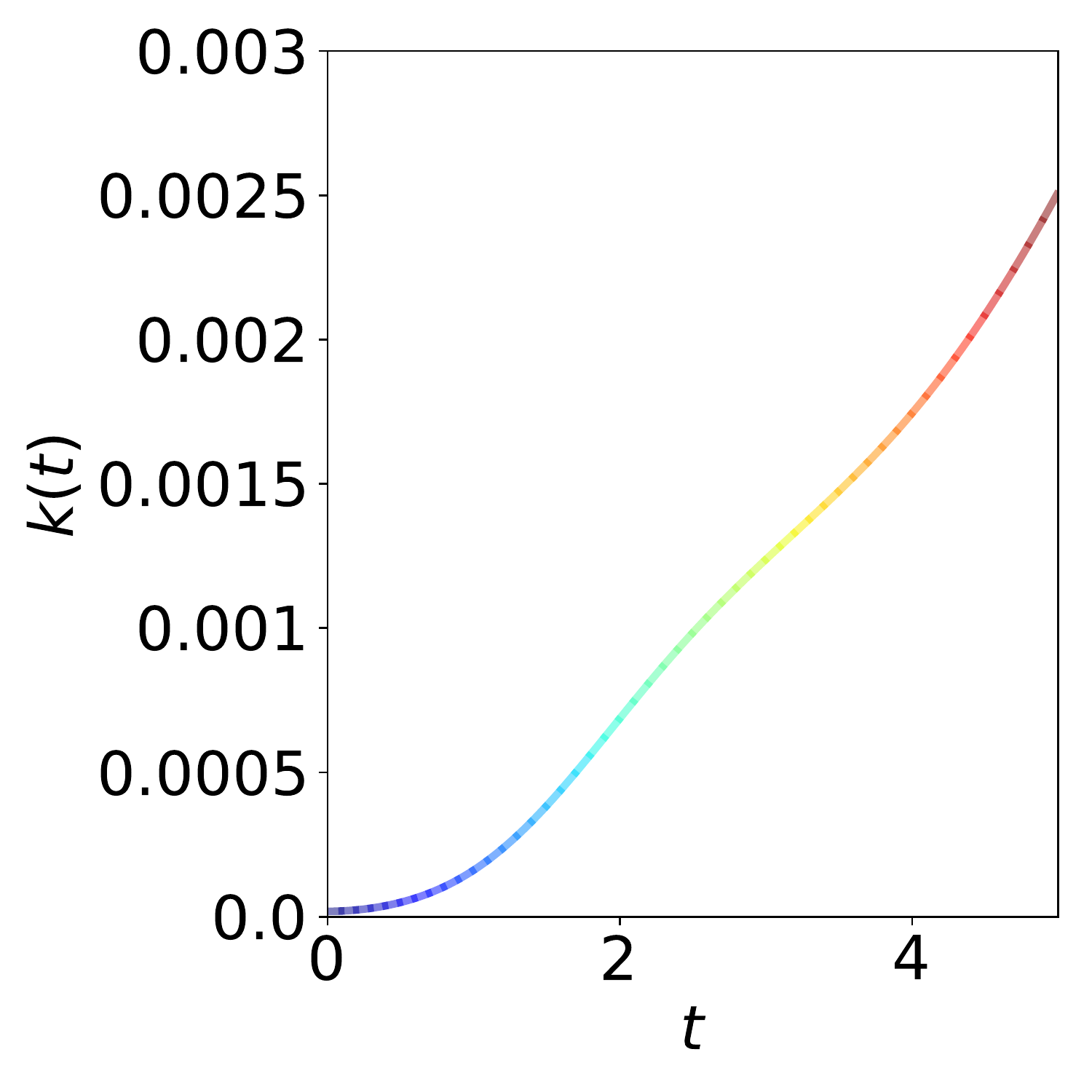}
    (c) \includegraphics[width=0.29 \textwidth]{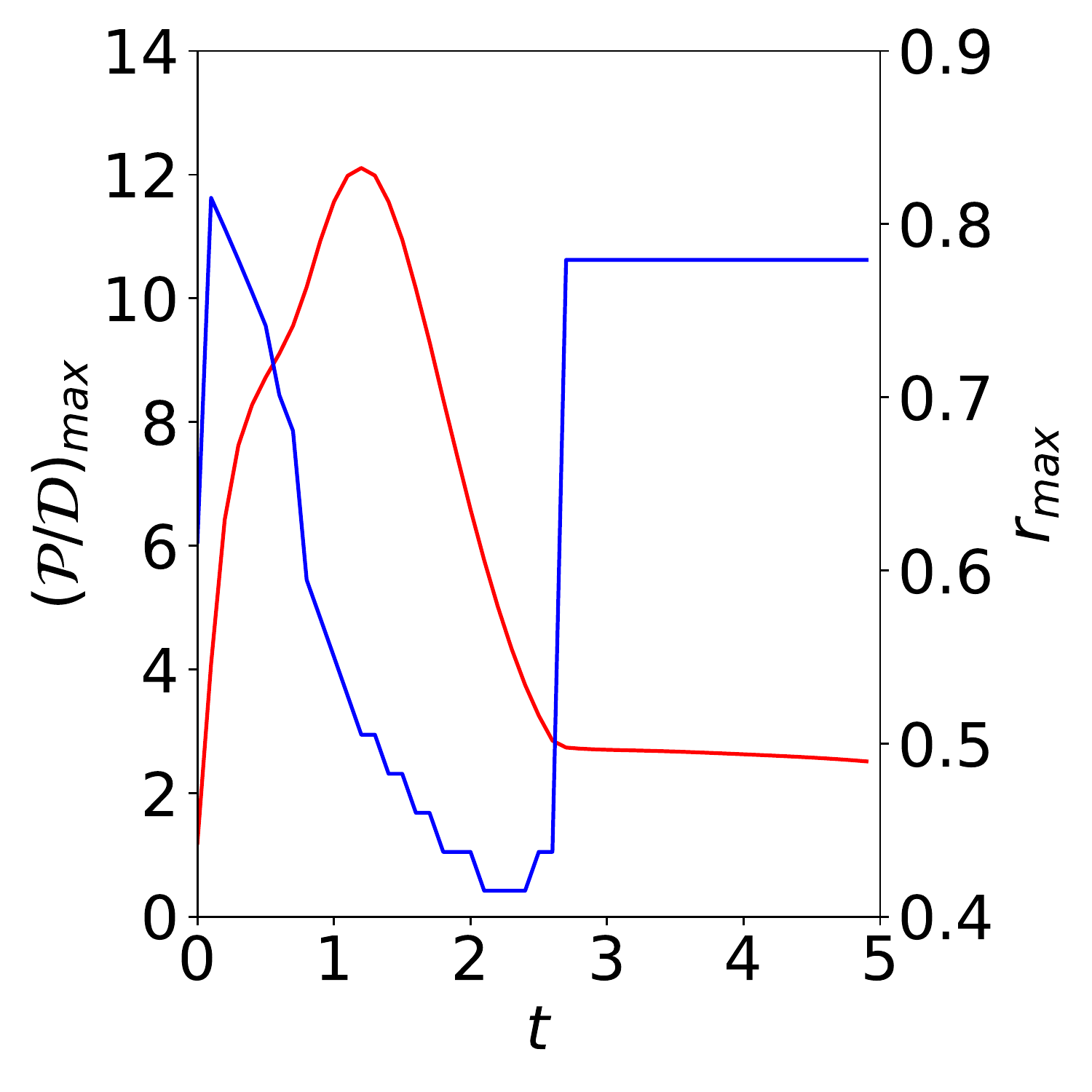}
	  \caption{
      (a) Time evolution of $[\product/\dissip] (r)$ for the minimal seed
	 at $Re$=3000. 
      Colors indicate the direction of time (initial:blue, final:red); 
      each curve is separated by $\Delta t = 0.1$ in time. 
      (b) Time-series of perturbation kinetic energy with the same color-coding for reference.
      (c) Time evolution of the peak $(\product/\dissip)_{max}$ (red) 
      and of its radial position $r_{max}$ (blue) for the minimal seed.
       \label{fig-PDtime-MS_t5}}  
\end{figure}
\begin{figure}
  \centering
    (a) \includegraphics[width=0.29\textwidth]{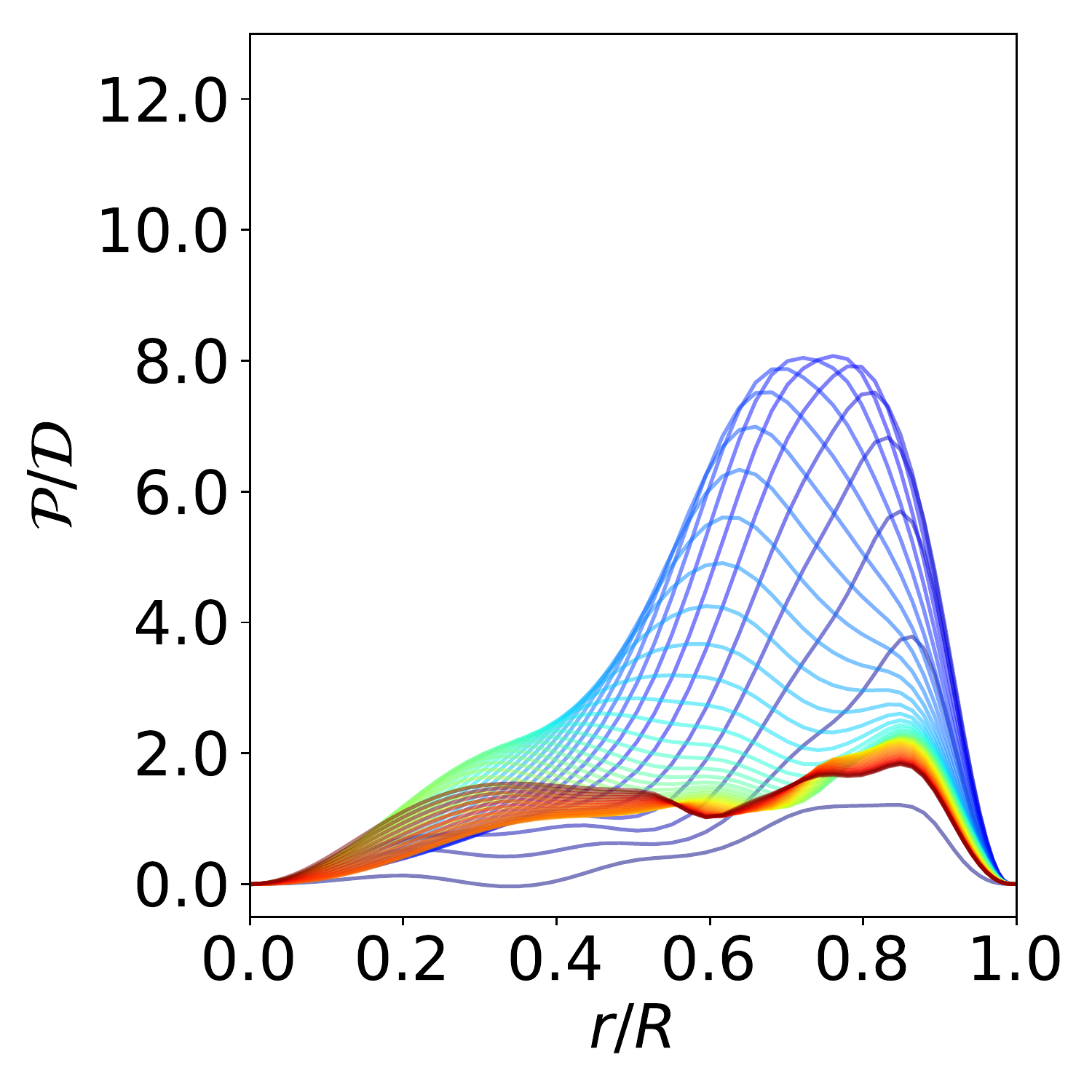}
	(b) \includegraphics[width=0.29\textwidth]{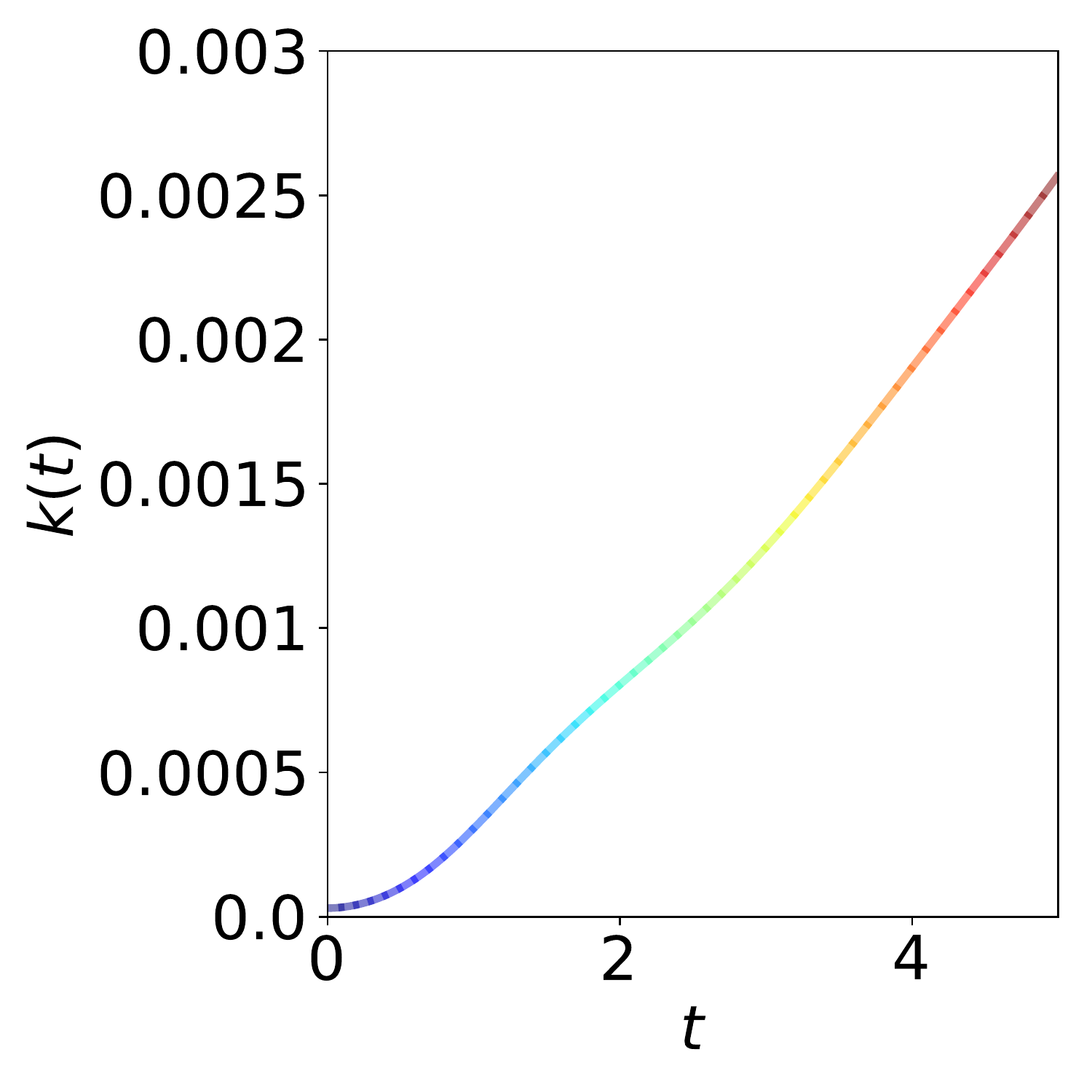}
    (c) \includegraphics[width=0.29 \textwidth]{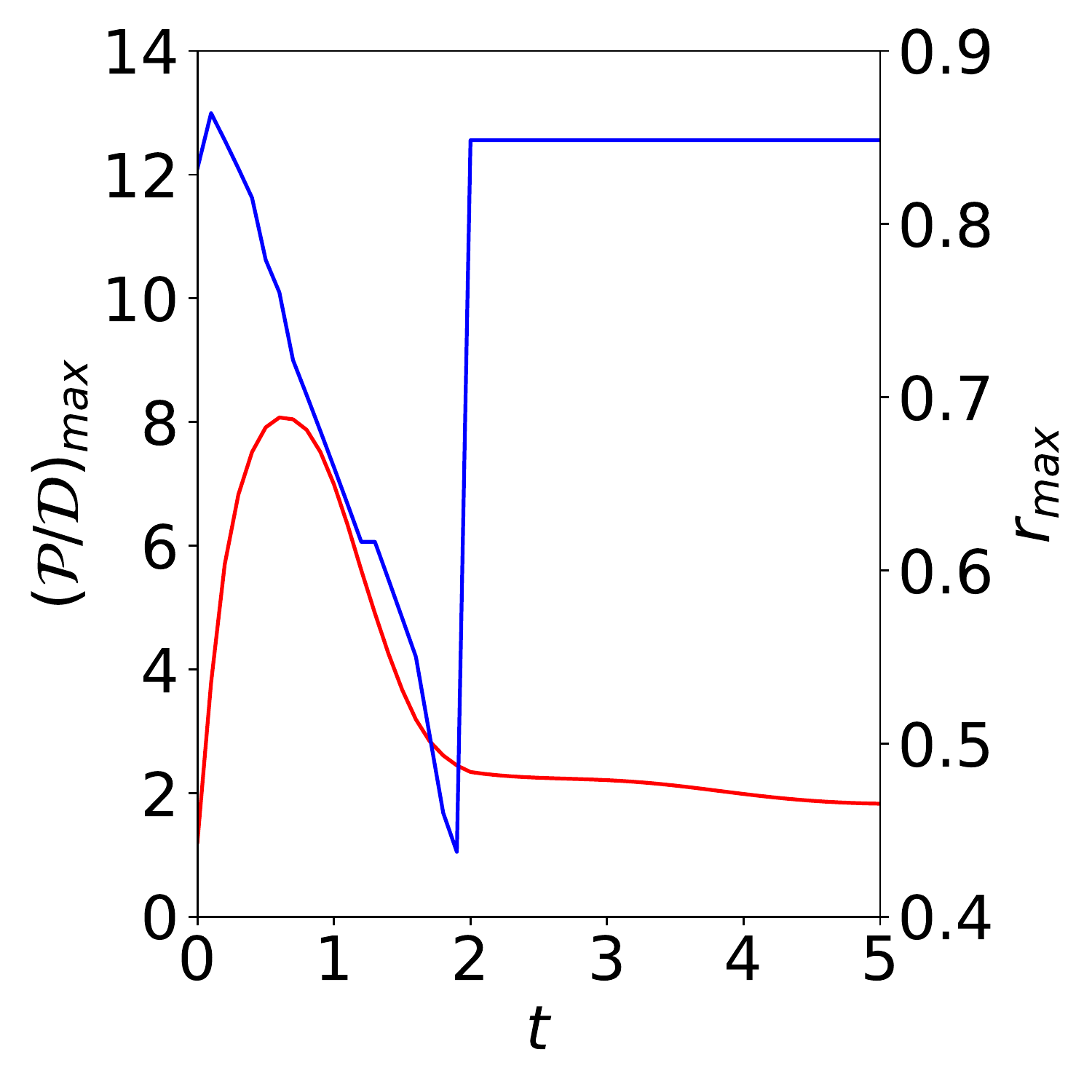}
	  \caption{
      (a) Time evolution of $[\product/\dissip] (r)$ for the minimal 
      seed at $Re$=3000 with flattened base profile (refer to 
      \eqref{forc-base-prof}).
       Colors indicate the direction of time (initial:blue, final:red); 
      each curve is separated by $\Delta t = 0.1$ in time.  
      (b) Time-series of perturbation kinetic energy with the 
      same color-coding for reference.
      (c) Time evolution of the peak $(\product/\dissip)_{max}$ (red) 
      and of its radial position $r_{max}$ (blue) for the minimal seed 
      with flattened base profile.
       \label{fig-PDtime-forced}}  
\end{figure}

Marensi \etal \rf{marensi-etal-2019} showed that flattening the base 
profile in a pipe flow not only destabilizes the turbulence, as in the 
experiment of K\"uhnen \etal \rf{kuhnen-etal-2018a}, but it also 
enhances the nonlinear stability of the laminar flow.
Minimal seed calculations were performed with a flattened base profile 
\citep{kuhnen-etal-2018a}, namely
  \beq
     U(r;\delta, \gamma)=
     (1-\delta)\left[1 - \frac{\cosh(\gamma r) - 1}{\cosh(\gamma) - 1} \right]\, ,
     \label{forc-base-prof}
  \eeq
where $\delta$ is the centerline difference between the laminar profile 
and the target profile and $\gamma \approx 2.424$ is set by the constant 
mass flux condition. The critical initial energy of the minimal seed was found to 
increase with increasing values of $\delta$, that is, the flattening 
expands the basin of attraction of the laminar state.
Here, we analyze this phenomenon in the light of the 
$[\product/\dissip](r)$ profiles and of their time evolution.
For example, an interesting question is the following: which of the 
phases of the minimal seed amplification is the flattening of the base 
profile affecting and how?
First, we repeat the minimal seed calculation of 
\refref{marensi-etal-2019} for $Re=3000$ and at $\delta=0.12$. The 
corresponding $[\product/\dissip](r)$ curves are shown in 
\reffig{fig-PDtime-forced}.
By comparing it with \reffig{fig-PDtime-MS_t5}, we observe that the 
production in the bulk region is strongly reduced by the flattening.
\refFig{fig-PDtime-forced} also shows that all three phases of the 
minimal seed amplification are still present in the forced case, but 
they are not able to achieve as much energy production as in the 
unforced case (for example in the Orr phase, the peak 
$(\product/\dissip)_{max}$ is almost $30\%$ lower than with the 
parabolic base profile.).
These observations illustrate that strategies to delay transition have
to take the central part of the flow into account. Since transition relies on 
the production away from the wall, a flatter velocity profile effectively 
reduces the susceptibility of the flow to perturbations.
\section{Conclusion}
\label{s-conclusion}

In this paper, we demonstrated that in pipe flow at the transitional stages,
velocity field fluctuations are amplified away from the wall 
$r \in (0.4R, 0.8R)$ as opposed to the typical turbulent fluctuations, which 
are predominantly generated near the wall $r \approx 0.9R$. To this end, 
we numerically investigated various transition scenarios including the 
perturbations that are too energetic to initiate turbulence 
(the upper edge) and the perturbations that are only energetic enough
(the minimal seeds) for the transition. In all cases, we observed that the
strong energy amplification away from the wall is an essential stage of 
transition.

     Although we focused here on the production and dissipation terms for
	 our diagnostics, it should be noted that the full energy balance 
	 equation of the pipe flow also have pressure, viscous, and nonlinear
	 transport terms\rf{popebook}. 
	 For the highly-turbulent initial conditions we studied 
	 in Section \ref{s-upper}, we found these terms to be negligible in the 
	 bulk region.
 	 For the traveling waves and other snapshots from the edge state, 
 	 while the pressure transport was still 
 	 very small, the viscous and nonlinear terms had peaks located close
 	 to that of the production. 
  	 This trend is also observed in the turbulent channel flow \rf{popebook}
  	 when these terms are evaluated as a function of the wall-normal 
  	 coordinate and interpreted as the redistribution of excess
  	 energy due to production peak.  
  	 We believe that our case here is similar since the edge state
  	 is dynamically invariant and, thus, it must exhibit energy 
  	 balance. 
   	 
    In both transitions from the edge state 
	(\reffig{f-prodispt}) and from the minimal seeds
	(figs. \ref{fig-PDtime-MS}--\ref{fig-PDtime-forced}), 
	kinetic energy of the velocity perturbations show episodes
	of strong amplification. 
	Notice that all of the corresponding $\product/\dissip$ 
	curves have values greater than $1$ in the bulk of the 
	domain. 
	This is expected since the volume integral of 
	$\product - \dissip$ gives the instantaneous change 
	in total kinetic energy of the velocity fluctuations.

During the transition from the minimal seed, 
the peak of $ [\product/\dissip] (r)$ and its radial location 
have proven to be a clear 
indicator of which 
state of transition is taking place.
Such knowledge is of great importance when investigating the nonlinear 
stability of shear flows in subcritical transition scenarios as it 
gives insights on which growth mechanism to target in order to enhance 
or suppress transition.

In many industrial applications, turbulence is undesirable due to its
high energy cost. Our results indicate that the control strategies,
which aim to avoid transition must eliminate fluctuations away 
from the pipe wall. 
Furthermore,
\refref{kuhnen-etal-2018a} measured transient growth due to the non-normality
of the linearized Navier-Stokes operator assuming the mean profile to be 
the base flow. They found that the transient growth was substantially 
suppressed when the profile was flattened.
 Our results suggest that the 
transient growth of the mean profile serves as a proxy of the production
away from the wall, which is necessary for the transition to turbulence.

\bibliography{../../../bibtex/pipes}

\begin{thebibliography}{29}%
\makeatletter
\providecommand \@ifxundefined [1]{%
 \@ifx{#1\undefined}
}%
\providecommand \@ifnum [1]{%
 \ifnum #1\expandafter \@firstoftwo
 \else \expandafter \@secondoftwo
 \fi
}%
\providecommand \@ifx [1]{%
 \ifx #1\expandafter \@firstoftwo
 \else \expandafter \@secondoftwo
 \fi
}%
\providecommand \natexlab [1]{#1}%
\providecommand \enquote  [1]{``#1''}%
\providecommand \bibnamefont  [1]{#1}%
\providecommand \bibfnamefont [1]{#1}%
\providecommand \citenamefont [1]{#1}%
\providecommand \href@noop [0]{\@secondoftwo}%
\providecommand \href [0]{\begingroup \@sanitize@url \@href}%
\providecommand \@href[1]{\@@startlink{#1}\@@href}%
\providecommand \@@href[1]{\endgroup#1\@@endlink}%
\providecommand \@sanitize@url [0]{\catcode `\\12\catcode `\$12\catcode
  `\&12\catcode `\#12\catcode `\^12\catcode `\_12\catcode `\%12\relax}%
\providecommand \@@startlink[1]{}%
\providecommand \@@endlink[0]{}%
\providecommand \url  [0]{\begingroup\@sanitize@url \@url }%
\providecommand \@url [1]{\endgroup\@href {#1}{\urlprefix }}%
\providecommand \urlprefix  [0]{URL }%
\providecommand \Eprint [0]{\href }%
\providecommand \doibase [0]{http://dx.doi.org/}%
\providecommand \selectlanguage [0]{\@gobble}%
\providecommand \bibinfo  [0]{\@secondoftwo}%
\providecommand \bibfield  [0]{\@secondoftwo}%
\providecommand \translation [1]{[#1]}%
\providecommand \BibitemOpen [0]{}%
\providecommand \bibitemStop [0]{}%
\providecommand \bibitemNoStop [0]{.\EOS\space}%
\providecommand \EOS [0]{\spacefactor3000\relax}%
\providecommand \BibitemShut  [1]{\csname bibitem#1\endcsname}%
\let\auto@bib@innerbib\@empty
\bibitem [{\citenamefont {Reynolds}(1883)}]{R1883}%
  \BibitemOpen
  \bibfield  {author} {\bibinfo {author} {\bibfnamefont {O.}~\bibnamefont
  {Reynolds}},\ }\bibfield  {title} {\enquote {\bibinfo {title} {An
  experimental investigation of the circumstances which determine whether the
  motion of water shall be direct or sinuous, and the law of resistance in
  parallel channels},}\ }\href@noop {} {\bibfield  {journal} {\bibinfo
  {journal} {Proc. Roy. Soc. Lond. Ser A}\ }\textbf {\bibinfo {volume} {174}},\
  \bibinfo {pages} {935--982} (\bibinfo {year} {1883})}\BibitemShut {NoStop}%
\bibitem [{\citenamefont {Skufca}\ \emph {et~al.}(2006)\citenamefont {Skufca},
  \citenamefont {Yorke},\ and\ \citenamefont {Eckhardt}}]{SYE05}%
  \BibitemOpen
  \bibfield  {author} {\bibinfo {author} {\bibfnamefont {J.~D.}\ \bibnamefont
  {Skufca}}, \bibinfo {author} {\bibfnamefont {J.~A.}\ \bibnamefont {Yorke}}, \
  and\ \bibinfo {author} {\bibfnamefont {B.}~\bibnamefont {Eckhardt}},\
  }\bibfield  {title} {\enquote {\bibinfo {title} {{Edge of Chaos} in a
  parallel shear flow},}\ }\href {\doibase 10.1103/PhysRevLett.96.174101}
  {\bibfield  {journal} {\bibinfo  {journal} {Phys. Rev. Lett.}\ }\textbf
  {\bibinfo {volume} {96}},\ \bibinfo {pages} {174101} (\bibinfo {year}
  {2006})}\BibitemShut {NoStop}%
\bibitem [{\citenamefont {Hopf}(1948)}]{hopf48}%
  \BibitemOpen
  \bibfield  {author} {\bibinfo {author} {\bibfnamefont {E.}~\bibnamefont
  {Hopf}},\ }\bibfield  {title} {\enquote {\bibinfo {title} {A mathematical
  example displaying features of turbulence},}\ }\href {\doibase
  10.1002/cpa.3160010401} {\bibfield  {journal} {\bibinfo  {journal} {Commun.
  Pure Appl. Math.}\ }\textbf {\bibinfo {volume} {1}},\ \bibinfo {pages}
  {303--322} (\bibinfo {year} {1948})}\BibitemShut {NoStop}%
\bibitem [{\citenamefont {Alligood}\ \emph {et~al.}(1997)\citenamefont
  {Alligood}, \citenamefont {Sauer},\ and\ \citenamefont
  {Yorke}}]{ASY1997ch10}%
  \BibitemOpen
  \bibfield  {author} {\bibinfo {author} {\bibfnamefont {K.~T.}\ \bibnamefont
  {Alligood}}, \bibinfo {author} {\bibfnamefont {T.~D.}\ \bibnamefont {Sauer}},
  \ and\ \bibinfo {author} {\bibfnamefont {J.~A.}\ \bibnamefont {Yorke}},\
  }\enquote {\bibinfo {title} {Stable manifolds and crises},}\ in\ \href
  {\doibase 10.1007/978-3-642-59281-2_10} {\emph {\bibinfo {booktitle} {Chaos:
  An Introduction to Dynamical Systems}}}\ (\bibinfo  {publisher} {Springer
  Berlin Heidelberg},\ \bibinfo {address} {Berlin, Heidelberg},\ \bibinfo
  {year} {1997})\ pp.\ \bibinfo {pages} {399--445}\BibitemShut {NoStop}%
\bibitem [{\citenamefont {Grebogi}\ \emph {et~al.}(1982)\citenamefont
  {Grebogi}, \citenamefont {Ott},\ and\ \citenamefont {Yorke}}]{GOY1982}%
  \BibitemOpen
  \bibfield  {author} {\bibinfo {author} {\bibfnamefont {C.}~\bibnamefont
  {Grebogi}}, \bibinfo {author} {\bibfnamefont {E.}~\bibnamefont {Ott}}, \ and\
  \bibinfo {author} {\bibfnamefont {J.~A.}\ \bibnamefont {Yorke}},\ }\bibfield
  {title} {\enquote {\bibinfo {title} {Chaotic attractors in crisis},}\ }\href
  {\doibase 10.1103/PhysRevLett.48.1507} {\bibfield  {journal} {\bibinfo
  {journal} {Phys. Rev. Lett.}\ }\textbf {\bibinfo {volume} {48}},\ \bibinfo
  {pages} {1507--1510} (\bibinfo {year} {1982})}\BibitemShut {NoStop}%
\bibitem [{\citenamefont {Grebogi}\ \emph {et~al.}(1983)\citenamefont
  {Grebogi}, \citenamefont {Ott},\ and\ \citenamefont {Yorke}}]{GOY1983}%
  \BibitemOpen
  \bibfield  {author} {\bibinfo {author} {\bibfnamefont {C.}~\bibnamefont
  {Grebogi}}, \bibinfo {author} {\bibfnamefont {E.}~\bibnamefont {Ott}}, \ and\
  \bibinfo {author} {\bibfnamefont {J.~A.}\ \bibnamefont {Yorke}},\ }\bibfield
  {title} {\enquote {\bibinfo {title} {Crises, sudden changes in chaotic
  attractors, and transient chaos},}\ }\href {\doibase
  10.1016/0167-2789(83)90126-4} {\bibfield  {journal} {\bibinfo  {journal}
  {Physica D}\ }\textbf {\bibinfo {volume} {7}},\ \bibinfo {pages} {181--200}
  (\bibinfo {year} {1983})}\BibitemShut {NoStop}%
\bibitem [{\citenamefont {Itano}\ and\ \citenamefont {Toh}(2001)}]{IT01}%
  \BibitemOpen
  \bibfield  {author} {\bibinfo {author} {\bibfnamefont {T.}~\bibnamefont
  {Itano}}\ and\ \bibinfo {author} {\bibfnamefont {S.}~\bibnamefont {Toh}},\
  }\bibfield  {title} {\enquote {\bibinfo {title} {The dynamics of bursting
  process in wall turbulence},}\ }\href@noop {} {\bibfield  {journal} {\bibinfo
   {journal} {J. Phys. Soc. Japan}\ }\textbf {\bibinfo {volume} {70}},\
  \bibinfo {pages} {701--714} (\bibinfo {year} {2001})}\BibitemShut {NoStop}%
\bibitem [{\citenamefont {Toh}\ and\ \citenamefont {Itano}(2003)}]{TI03}%
  \BibitemOpen
  \bibfield  {author} {\bibinfo {author} {\bibfnamefont {S.}~\bibnamefont
  {Toh}}\ and\ \bibinfo {author} {\bibfnamefont {T.}~\bibnamefont {Itano}},\
  }\bibfield  {title} {\enquote {\bibinfo {title} {A periodic-like solution in
  channel flow},}\ }\href@noop {} {\bibfield  {journal} {\bibinfo  {journal}
  {J. Fluid Mech.}\ }\textbf {\bibinfo {volume} {481}},\ \bibinfo {pages}
  {67--76} (\bibinfo {year} {2003})}\BibitemShut {NoStop}%
\bibitem [{\citenamefont {Schneider}\ \emph {et~al.}(2007)\citenamefont
  {Schneider}, \citenamefont {Eckhardt},\ and\ \citenamefont
  {Yorke}}]{SchEckYor07}%
  \BibitemOpen
  \bibfield  {author} {\bibinfo {author} {\bibfnamefont {T.~M.}\ \bibnamefont
  {Schneider}}, \bibinfo {author} {\bibfnamefont {B.}~\bibnamefont {Eckhardt}},
  \ and\ \bibinfo {author} {\bibfnamefont {J.}~\bibnamefont {Yorke}},\
  }\bibfield  {title} {\enquote {\bibinfo {title} {Turbulence, transition, and
  the edge of chaos in pipe flow},}\ }\href {\doibase
  10.1103/PhysRevLett.99.034502} {\bibfield  {journal} {\bibinfo  {journal}
  {Phys. Rev. Lett.}\ }\textbf {\bibinfo {volume} {99}},\ \bibinfo {pages}
  {034502} (\bibinfo {year} {2007})}\BibitemShut {NoStop}%
\bibitem [{\citenamefont {Schneider}\ \emph {et~al.}(2008)\citenamefont
  {Schneider}, \citenamefont {Gibson}, \citenamefont {Lagha}, \citenamefont
  {Lillo},\ and\ \citenamefont {Eckhardt}}]{SGLDE08}%
  \BibitemOpen
  \bibfield  {author} {\bibinfo {author} {\bibfnamefont {T.~M.}\ \bibnamefont
  {Schneider}}, \bibinfo {author} {\bibfnamefont {J.~F.}\ \bibnamefont
  {Gibson}}, \bibinfo {author} {\bibfnamefont {M.}~\bibnamefont {Lagha}},
  \bibinfo {author} {\bibfnamefont {F.~De}\ \bibnamefont {Lillo}}, \ and\
  \bibinfo {author} {\bibfnamefont {B.}~\bibnamefont {Eckhardt}},\ }\bibfield
  {title} {\enquote {\bibinfo {title} {Laminar-turbulent boundary in plane
  {Couette} flow},}\ }\href@noop {} {\bibfield  {journal} {\bibinfo  {journal}
  {Phys. Rev. E.}\ }\textbf {\bibinfo {volume} {78}},\ \bibinfo {pages}
  {037301} (\bibinfo {year} {2008})},\ \bibinfo {note}
  {\arXiv{0805.1015}}\BibitemShut {NoStop}%
\bibitem [{\citenamefont {Mellibovsky}\ \emph {et~al.}(2009)\citenamefont
  {Mellibovsky}, \citenamefont {Meseguer}, \citenamefont {Schneider},\ and\
  \citenamefont {Eckhardt}}]{MMSE09}%
  \BibitemOpen
  \bibfield  {author} {\bibinfo {author} {\bibfnamefont {F.}~\bibnamefont
  {Mellibovsky}}, \bibinfo {author} {\bibfnamefont {A.}~\bibnamefont
  {Meseguer}}, \bibinfo {author} {\bibfnamefont {T.~M.}\ \bibnamefont
  {Schneider}}, \ and\ \bibinfo {author} {\bibfnamefont {B.}~\bibnamefont
  {Eckhardt}},\ }\bibfield  {title} {\enquote {\bibinfo {title} {Transition in
  localized pipe flow turbulence},}\ }\href {\doibase
  10.1103/PhysRevLett.103.054502} {\bibfield  {journal} {\bibinfo  {journal}
  {Phys. Rev. Lett.}\ }\textbf {\bibinfo {volume} {103}},\ \bibinfo {pages}
  {054502} (\bibinfo {year} {2009})}\BibitemShut {NoStop}%
\bibitem [{\citenamefont {Schneider}\ \emph {et~al.}(2010)\citenamefont
  {Schneider}, \citenamefont {Marinc},\ and\ \citenamefont
  {Eckhardt}}]{ScMaEc10}%
  \BibitemOpen
  \bibfield  {author} {\bibinfo {author} {\bibfnamefont {T.~M.}\ \bibnamefont
  {Schneider}}, \bibinfo {author} {\bibfnamefont {D.}~\bibnamefont {Marinc}}, \
  and\ \bibinfo {author} {\bibfnamefont {B.}~\bibnamefont {Eckhardt}},\
  }\bibfield  {title} {\enquote {\bibinfo {title} {Localized edge states
  nucleate turbulence in extended plane {Couette} cells},}\ }\href {\doibase
  10.1017/S0022112009993144} {\bibfield  {journal} {\bibinfo  {journal} {J.
  Fluid Mech.}\ }\textbf {\bibinfo {volume} {646}},\ \bibinfo {pages}
  {441--451} (\bibinfo {year} {2010})}\BibitemShut {NoStop}%
\bibitem [{\citenamefont {Zammert}\ and\ \citenamefont
  {Eckhardt}(2014)}]{ZamEck14a}%
  \BibitemOpen
  \bibfield  {author} {\bibinfo {author} {\bibfnamefont {S.}~\bibnamefont
  {Zammert}}\ and\ \bibinfo {author} {\bibfnamefont {B.}~\bibnamefont
  {Eckhardt}},\ }\bibfield  {title} {\enquote {\bibinfo {title} {A spotlike
  edge state in plane {Poiseuille} flow},}\ }\href {\doibase
  10.1002/pamm.201410283} {\bibfield  {journal} {\bibinfo  {journal} {PAMM}\
  }\textbf {\bibinfo {volume} {14}},\ \bibinfo {pages} {591--592} (\bibinfo
  {year} {2014})}\BibitemShut {NoStop}%
\bibitem [{\citenamefont {Khapko}\ \emph {et~al.}(2016)\citenamefont {Khapko},
  \citenamefont {Kreilos}, \citenamefont {Schlatter}, \citenamefont {Duguet},
  \citenamefont {Eckhardt},\ and\ \citenamefont {Henningson}}]{KKSDEH2016}%
  \BibitemOpen
  \bibfield  {author} {\bibinfo {author} {\bibfnamefont {T.}~\bibnamefont
  {Khapko}}, \bibinfo {author} {\bibfnamefont {T.}~\bibnamefont {Kreilos}},
  \bibinfo {author} {\bibfnamefont {P.}~\bibnamefont {Schlatter}}, \bibinfo
  {author} {\bibfnamefont {Y.}~\bibnamefont {Duguet}}, \bibinfo {author}
  {\bibfnamefont {B.}~\bibnamefont {Eckhardt}}, \ and\ \bibinfo {author}
  {\bibfnamefont {D.~S.}\ \bibnamefont {Henningson}},\ }\bibfield  {title}
  {\enquote {\bibinfo {title} {Edge states as mediators of bypass transition in
  boundary-layer flows},}\ }\href {\doibase 10.1017/jfm.2016.434} {\bibfield
  {journal} {\bibinfo  {journal} {J. Fluid Mech.}\ }\textbf {\bibinfo {volume}
  {801}} (\bibinfo {year} {2016}),\ 10.1017/jfm.2016.434}\BibitemShut {NoStop}%
\bibitem [{\citenamefont {Budanur}\ and\ \citenamefont
  {Hof}(2018)}]{budanur-hof-2018}%
  \BibitemOpen
  \bibfield  {author} {\bibinfo {author} {\bibfnamefont {N.~B.}\ \bibnamefont
  {Budanur}}\ and\ \bibinfo {author} {\bibfnamefont {B.}~\bibnamefont {Hof}},\
  }\bibfield  {title} {\enquote {\bibinfo {title} {Complexity of the
  laminar-turbulent boundary in pipe flow},}\ }\href {\doibase
  10.1103/PhysRevFluids.3.054401} {\bibfield  {journal} {\bibinfo  {journal}
  {Phys. Rev. Fluids}\ }\textbf {\bibinfo {volume} {3}},\ \bibinfo {pages}
  {054401} (\bibinfo {year} {2018})}\BibitemShut {NoStop}%
\bibitem [{\citenamefont {Pringle}\ and\ \citenamefont
  {Kerswell}(2010)}]{pringle-kerswell-2010}%
  \BibitemOpen
  \bibfield  {author} {\bibinfo {author} {\bibfnamefont {C.~C.~T.}\
  \bibnamefont {Pringle}}\ and\ \bibinfo {author} {\bibfnamefont {R.~R.}\
  \bibnamefont {Kerswell}},\ }\bibfield  {title} {\enquote {\bibinfo {title}
  {Using nonlinear transient growth to construct the minimal seed for shear
  flow turbulence},}\ }\href {\doibase 10.1103/PhysRevLett.105.154502}
  {\bibfield  {journal} {\bibinfo  {journal} {Phys. Rev. Lett.}\ }\textbf
  {\bibinfo {volume} {105}},\ \bibinfo {pages} {154502} (\bibinfo {year}
  {2010})}\BibitemShut {NoStop}%
\bibitem [{\citenamefont {Pringle}\ \emph {et~al.}(2012)\citenamefont
  {Pringle}, \citenamefont {Willis},\ and\ \citenamefont
  {Kerswell}}]{pringle-etal-2012}%
  \BibitemOpen
  \bibfield  {author} {\bibinfo {author} {\bibfnamefont {C.~C.~T.}\
  \bibnamefont {Pringle}}, \bibinfo {author} {\bibfnamefont {A.~P.}\
  \bibnamefont {Willis}}, \ and\ \bibinfo {author} {\bibfnamefont {R.~R.}\
  \bibnamefont {Kerswell}},\ }\bibfield  {title} {\enquote {\bibinfo {title}
  {Minimal seeds for shear flow turbulence: using nonlinear transient growth to
  touch the edge of chaos},}\ }\href@noop {} {\bibfield  {journal} {\bibinfo
  {journal} {J. Fluid Mech.}\ }\textbf {\bibinfo {volume} {702}},\ \bibinfo
  {pages} {415--443} (\bibinfo {year} {2012})}\BibitemShut {NoStop}%
\bibitem [{\citenamefont {Duguet}\ \emph {et~al.}(2013)\citenamefont {Duguet},
  \citenamefont {Monokrousos}, \citenamefont {Brandt},\ and\ \citenamefont
  {Henningson}}]{duguet-etal-2013}%
  \BibitemOpen
  \bibfield  {author} {\bibinfo {author} {\bibfnamefont {Y.}~\bibnamefont
  {Duguet}}, \bibinfo {author} {\bibfnamefont {A.}~\bibnamefont {Monokrousos}},
  \bibinfo {author} {\bibfnamefont {L.}~\bibnamefont {Brandt}}, \ and\ \bibinfo
  {author} {\bibfnamefont {D.~S.}\ \bibnamefont {Henningson}},\ }\bibfield
  {title} {\enquote {\bibinfo {title} {Minimal transition thresholds in plane
  {C}ouette flow},}\ }\href@noop {} {\bibfield  {journal} {\bibinfo  {journal}
  {Physics of Fluids}\ }\textbf {\bibinfo {volume} {25}},\ \bibinfo {pages}
  {084103} (\bibinfo {year} {2013})}\BibitemShut {NoStop}%
\bibitem [{\citenamefont {Cherubini}\ and\ \citenamefont
  {Palma}(2014)}]{cherubini-palma-2014}%
  \BibitemOpen
  \bibfield  {author} {\bibinfo {author} {\bibfnamefont {S.}~\bibnamefont
  {Cherubini}}\ and\ \bibinfo {author} {\bibfnamefont {P.~De}\ \bibnamefont
  {Palma}},\ }\bibfield  {title} {\enquote {\bibinfo {title} {Minimal
  perturbations approaching the edge of chaos in a couette flow},}\ }\href
  {http://stacks.iop.org/1873-7005/46/i=4/a=041403} {\bibfield  {journal}
  {\bibinfo  {journal} {Fluid Dyn. Res}\ }\textbf {\bibinfo {volume} {46}},\
  \bibinfo {pages} {041403} (\bibinfo {year} {2014})}\BibitemShut {NoStop}%
\bibitem [{\citenamefont {Kerswell}\ \emph {et~al.}(2014)\citenamefont
  {Kerswell}, \citenamefont {Pringle},\ and\ \citenamefont
  {Willis}}]{kerswell-etal-2014}%
  \BibitemOpen
  \bibfield  {author} {\bibinfo {author} {\bibfnamefont {R.~R.}\ \bibnamefont
  {Kerswell}}, \bibinfo {author} {\bibfnamefont {C.~C.~T.}\ \bibnamefont
  {Pringle}}, \ and\ \bibinfo {author} {\bibfnamefont {A.~P.}\ \bibnamefont
  {Willis}},\ }\bibfield  {title} {\enquote {\bibinfo {title} {An optimization
  approach for analysing nonlinear stability with transition to turbulence in
  fluids as an exemplar},}\ }\href@noop {} {\bibfield  {journal} {\bibinfo
  {journal} {Rep. Progr. Phys.}\ }\textbf {\bibinfo {volume} {77}},\ \bibinfo
  {pages} {085901} (\bibinfo {year} {2014})}\BibitemShut {NoStop}%
\bibitem [{\citenamefont {Marensi}\ \emph {et~al.}(2019)\citenamefont
  {Marensi}, \citenamefont {Willis},\ and\ \citenamefont
  {Kerswell}}]{marensi-etal-2019}%
  \BibitemOpen
  \bibfield  {author} {\bibinfo {author} {\bibfnamefont {E.}~\bibnamefont
  {Marensi}}, \bibinfo {author} {\bibfnamefont {A.~P.}\ \bibnamefont {Willis}},
  \ and\ \bibinfo {author} {\bibfnamefont {R.~R.}\ \bibnamefont {Kerswell}},\
  }\bibfield  {title} {\enquote {\bibinfo {title} {Stabilisation and drag
  reduction of pipe flows by flattening the base profile},}\ }\href@noop {}
  {\bibfield  {journal} {\bibinfo  {journal} {J. Fluid Mech.}\ }\textbf
  {\bibinfo {volume} {863}},\ \bibinfo {pages} {850--875} (\bibinfo {year}
  {2019})}\BibitemShut {NoStop}%
\bibitem [{\citenamefont {Joseph}\ and\ \citenamefont
  {Hung}(1971)}]{Joseph1971}%
  \BibitemOpen
  \bibfield  {author} {\bibinfo {author} {\bibfnamefont {D.~D.}\ \bibnamefont
  {Joseph}}\ and\ \bibinfo {author} {\bibfnamefont {W.}~\bibnamefont {Hung}},\
  }\bibfield  {title} {\enquote {\bibinfo {title} {Contributions to the
  nonlinear theory of stability of viscous flow in pipes and between rotating
  cylinders},}\ }\href {\doibase 10.1007/BF00250825} {\bibfield  {journal}
  {\bibinfo  {journal} {Arch. Rational Mech. Anal.}\ }\textbf {\bibinfo
  {volume} {44}},\ \bibinfo {pages} {1--22} (\bibinfo {year}
  {1971})}\BibitemShut {NoStop}%
\bibitem [{\citenamefont {K\"{u}hnen}\ \emph {et~al.}(2018)\citenamefont
  {K\"{u}hnen}, \citenamefont {Song}, \citenamefont {Scarselli}, \citenamefont
  {Budanur}, \citenamefont {Riedl}, \citenamefont {Willis}, \citenamefont
  {Avila},\ and\ \citenamefont {Hof}}]{kuhnen-etal-2018a}%
  \BibitemOpen
  \bibfield  {author} {\bibinfo {author} {\bibfnamefont {J.}~\bibnamefont
  {K\"{u}hnen}}, \bibinfo {author} {\bibfnamefont {B.}~\bibnamefont {Song}},
  \bibinfo {author} {\bibfnamefont {D.}~\bibnamefont {Scarselli}}, \bibinfo
  {author} {\bibfnamefont {N.~B.}\ \bibnamefont {Budanur}}, \bibinfo {author}
  {\bibfnamefont {M.}~\bibnamefont {Riedl}}, \bibinfo {author} {\bibfnamefont
  {A.~P.}\ \bibnamefont {Willis}}, \bibinfo {author} {\bibfnamefont
  {M.}~\bibnamefont {Avila}}, \ and\ \bibinfo {author} {\bibfnamefont
  {B.}~\bibnamefont {Hof}},\ }\bibfield  {title} {\enquote {\bibinfo {title}
  {Destabilizing turbulence in pipe flow},}\ }\href {\doibase
  10.1038/s41567-017-0018-3} {\bibfield  {journal} {\bibinfo  {journal} {Nat.
  Phys.}\ }\textbf {\bibinfo {volume} {14}},\ \bibinfo {pages} {386--390}
  (\bibinfo {year} {2018})}\BibitemShut {NoStop}%
\bibitem [{\citenamefont {Willis}(2017)}]{willis-2017}%
  \BibitemOpen
  \bibfield  {author} {\bibinfo {author} {\bibfnamefont {A.~P.}\ \bibnamefont
  {Willis}},\ }\bibfield  {title} {\enquote {\bibinfo {title} {The
  {O}penpipeflow {N}avier--{S}tokes solver},}\ }\href@noop {} {\bibfield
  {journal} {\bibinfo  {journal} {SoftwareX}\ }\textbf {\bibinfo {volume}
  {6}},\ \bibinfo {pages} {124--127} (\bibinfo {year} {2017})}\BibitemShut
  {NoStop}%
\bibitem [{\citenamefont {Pope}(2000)}]{popebook}%
  \BibitemOpen
  \bibfield  {author} {\bibinfo {author} {\bibfnamefont {S.~B.}\ \bibnamefont
  {Pope}},\ }\href@noop {} {\emph {\bibinfo {title} {Turbulent Flows}}}\
  (\bibinfo  {publisher} {Cambridge Univ. Press},\ \bibinfo {address}
  {Cambridge},\ \bibinfo {year} {2000})\BibitemShut {NoStop}%
\bibitem [{\citenamefont {Kim}\ \emph {et~al.}(1987)\citenamefont {Kim},
  \citenamefont {Moin},\ and\ \citenamefont {Moser}}]{KMM87}%
  \BibitemOpen
  \bibfield  {author} {\bibinfo {author} {\bibfnamefont {J.}~\bibnamefont
  {Kim}}, \bibinfo {author} {\bibfnamefont {P.}~\bibnamefont {Moin}}, \ and\
  \bibinfo {author} {\bibfnamefont {R.}~\bibnamefont {Moser}},\ }\bibfield
  {title} {\enquote {\bibinfo {title} {Turbulence statistics in fully developed
  channel flow at low {Reynolds} number},}\ }\href@noop {} {\bibfield
  {journal} {\bibinfo  {journal} {J. Fluid Mech.}\ }\textbf {\bibinfo {volume}
  {177}},\ \bibinfo {pages} {133--166} (\bibinfo {year} {1987})}\BibitemShut
  {NoStop}%
\bibitem [{\citenamefont {Yang}\ \emph {et~al.}(2018)\citenamefont {Yang},
  \citenamefont {Willis},\ and\ \citenamefont {Hwang}}]{YWH2018}%
  \BibitemOpen
  \bibfield  {author} {\bibinfo {author} {\bibfnamefont {Q.}~\bibnamefont
  {Yang}}, \bibinfo {author} {\bibfnamefont {A.~P.}\ \bibnamefont {Willis}}, \
  and\ \bibinfo {author} {\bibfnamefont {Y.}~\bibnamefont {Hwang}},\ }\bibfield
   {title} {\enquote {\bibinfo {title} {Energy production and self-sustained
  turbulence at the kolmogorov scale in couette flow},}\ }\href {\doibase
  10.1017/jfm.2017.704} {\bibfield  {journal} {\bibinfo  {journal} {J. Fluid
  Mech.}\ }\textbf {\bibinfo {volume} {834}},\ \bibinfo {pages} {531--554}
  (\bibinfo {year} {2018})}\BibitemShut {NoStop}%
\bibitem [{\citenamefont {Avila}\ \emph {et~al.}(2013)\citenamefont {Avila},
  \citenamefont {Mellibovsky}, \citenamefont {Roland},\ and\ \citenamefont
  {Hof}}]{AvMeRoHo13}%
  \BibitemOpen
  \bibfield  {author} {\bibinfo {author} {\bibfnamefont {M.}~\bibnamefont
  {Avila}}, \bibinfo {author} {\bibfnamefont {F.}~\bibnamefont {Mellibovsky}},
  \bibinfo {author} {\bibfnamefont {N.}~\bibnamefont {Roland}}, \ and\ \bibinfo
  {author} {\bibfnamefont {B.}~\bibnamefont {Hof}},\ }\bibfield  {title}
  {\enquote {\bibinfo {title} {Streamwise-localized solutions at the onset of
  turbulence in pipe flow},}\ }\href {\doibase 10.1103/PhysRevLett.110.224502}
  {\bibfield  {journal} {\bibinfo  {journal} {Phys. Rev. Lett.}\ }\textbf
  {\bibinfo {volume} {110}},\ \bibinfo {pages} {224502} (\bibinfo {year}
  {2013})}\BibitemShut {NoStop}%
\bibitem [{\citenamefont {Duguet}\ \emph {et~al.}(2008)\citenamefont {Duguet},
  \citenamefont {Willis},\ and\ \citenamefont {Kerswell}}]{duguet07}%
  \BibitemOpen
  \bibfield  {author} {\bibinfo {author} {\bibfnamefont {Y.}~\bibnamefont
  {Duguet}}, \bibinfo {author} {\bibfnamefont {A.~P.}\ \bibnamefont {Willis}},
  \ and\ \bibinfo {author} {\bibfnamefont {R.~R.}\ \bibnamefont {Kerswell}},\
  }\bibfield  {title} {\enquote {\bibinfo {title} {Transition in pipe flow: the
  saddle structure on the boundary of turbulence},}\ }\href {\doibase
  10.1017/S0022112008003248} {\bibfield  {journal} {\bibinfo  {journal} {J.
  Fluid Mech.}\ }\textbf {\bibinfo {volume} {613}},\ \bibinfo {pages}
  {255--274} (\bibinfo {year} {2008})},\ \bibinfo {note}
  {\arXiv{0711.2175}}\BibitemShut {NoStop}%
\end{thebibliography}%
\end{document}